\documentclass[fleqn,usenatbib]{mnras}

\usepackage[T1]{fontenc}
\usepackage{ae,aecompl}

\usepackage{graphicx}	% Including figure files
\usepackage{amsmath}	% Advanced maths commands
\usepackage{amssymb}	% Extra maths symbols

\usepackage{verbatim}
\usepackage[autostyle]{csquotes}
\def\spose#1{\hbox to 0pt{#1\hss}}
\def\lta{\mathrel{\spose{\lower 3pt\hbox{$\mathchar"218$}}
     \raise 2.0pt\hbox{$\mathchar"13C$}}}
\def\gta{\mathrel{\spose{\lower 3pt\hbox{$\mathchar"218$}}
     \raise 2.0pt\hbox{$\mathchar"13E$}}}

\usepackage{newtxtext,newtxmath}
\title[Radio loudness of quasars from LoTSS]{The radio loudness of
  SDSS quasars from the LOFAR Two-metre Sky Survey: ubiquitous jet activity and constraints on star formation}

\author[C. Macfarlane et al.]{C. Macfarlane$^{1}$,
P. N. Best$^{1}$\thanks{E-mail: pnb@roe.ac.uk},
J. Sabater$^{1,2}$,
G. G\"{u}rkan$^{3,4}$,
M. J. Jarvis$^{5,6}$,
H. J. A. R{\"o}ttgering$^{7}$,
\newauthor
R. D. Baldi$^{8,9}$,
G. Calistro Rivera$^{10}$, 
K. J. Duncan$^{1,7}$,
L. K. Morabito$^{11}$,
I. Prandoni$^{8}$,
\newauthor
E. Retana-Montenegro$^{12}$
\\
$^{1}$Institute for Astronomy, University of Edinburgh, Royal Observatory, Blackford Hill, Edinburgh, EH9 3HJ, UK\\
$^{2}$UK Astronomy Technology Centre, Royal Observatory, Blackford Hill, Edinburgh, EH9 3HJ, UK\\
$^{3}$Th\"uringer Landessternwarte, Sternwarte 5, D-07778 Tautenburg, Germany\\
$^{4}$CSIRO Astronomy and Space Science, PO Box 1130, Bentley WA 6102, Australia\\
$^{5}$Astrophysics, Department of Physics, Keble Road, Oxford, OX1 3RH, UK\\
$^{6}$Department of Physics \& Astronomy, University of the Western Cape, Private Bag X17, Bellville, Cape Town, 7535, South Africa\\
$^{7}$Leiden Observatory, Leiden University, PO Box 9513, 2300 RA Leiden, The Netherlands\\ 
$^{8}$INAF - Istituto di Radioastronomia, Via Piero Gobetti 101, I-40129 Bologna, Italy\\
$^{9}$Department of Physics \& Astronomy, University of Southampton, Hampshire SO17 1BJ, Southampton, United Kingdom\\
$^{10}$European Southern Observatory, Karl-Schwarzschild-Stra{\ss}e 2, 85748 Garching bei M{\"u}nchen\\
$^{11}$Centre for Extragalactic Astronomy, Department of Physics, Durham University, Durham, DH1 3LE, UK\\ 
$^{12}$Astrophysics \& Cosmology Research Unit, School of Mathematics, Statistics \& Computer Science, University of KwaZulu-Natal, Durban 4041, South Africa
}

\date{Accepted XXX. Received YYY; in original form ZZZ}

\pubyear{2021}

\begin{document}
\label{firstpage}
\pagerange{\pageref{firstpage}--\pageref{lastpage}}
\maketitle

\begin{abstract}

We examine the distribution of radio emission from $\sim 42,000$
quasars from the Sloan Digital Sky Survey, as measured in the LOFAR
Two-Metre Sky Survey (LoTSS). We present a model of the radio
luminosity distribution of the quasars that assumes that every quasar
displays a superposition of two sources of radio emission: active
galactic nuclei (jets) and star-formation. Our two-component model
provides an excellent match to the observed radio flux density
distributions across a wide range of redshifts and quasar optical
luminosities; this suggests that the jet-launching mechanism operates
in all quasars but with different powering efficiency. The wide
distribution of jet powers allows for a smooth transition between the
`radio-quiet' and `radio-loud' quasar regimes, without need for any
explicit bimodality. The best-fit model parameters indicate that the
star-formation rate of quasar host galaxies correlates strongly with
quasar luminosity and also increases with redshift at least out to $z
\sim 2$. For a model where star-formation rate scales as $L_{\rm
  bol}^{\alpha} (1+z)^{\beta}$, we find $\alpha = 0.47 \pm 0.01$ and
$\beta = 1.61 \pm 0.05$, in agreement with far-infrared
studies. Quasars contribute $\approx 0.15$ per cent of the cosmic
star-formation rate density at $z=0.5$, rising to 0.4 per cent by $z
\sim 2$. The typical radio jet power is seen to increase with both
increasing optical luminosity and black hole mass independently, but
does not vary with redshift, suggesting intrinsic properties govern
the production of the radio jets.  We discuss the implications of
these results for the triggering of quasar activity and the launching
of jets.

% 250 words max (200 for letters)
\end{abstract}


%%%%%%%%%%%%%%%%%%%%%%%%%%%%%%%%%%%%%%%%%%%%%%%%%%

%%%%%%%%%%%%%%%%% BODY OF PAPER %%%%%%%%%%%%%%%%%%

\section{Introduction}
\label{Sec:intro}

The fundamental physical mechanism that powers and defines active
galactic nuclei (AGN) is the transfer of energy from the
relativistically-deep potential well of a central supermassive black
hole (SMBH) in a galaxy \citep[][]{Salpeter64, Zeldovich64, LB69}. The
most luminous members of the AGN class are quasars, which can outshine
their host galaxy on the order of hundreds or even thousands of
times. Traditionally, the ratio of the radio flux density to the
optical flux density of quasars has been used in the literature to
divide quasars into two categories: radio-loud (RL) and radio-quiet
\citep[RQ; e.g.][]{Kellermann1989}. By this definition, of the order
of 10\% of quasars are observed to be RL, although this fraction is
known to increase with both optical \citep{Padovani1993} and X-ray
\citep{DellaCeca1994} luminosity. Nevertheless, even RQ quasars have
been shown to emit weak radio emission when sufficiently deep radio
data are available \citep[e.g.][and references therein]{Padovani2016}.

Despite quasars being identified over half a century ago
\citep{Schmidt63}, the question of whether RL and RQ quasars are two
physically distinct populations has still not been answered. A
dichotomy appears to exist in the sense that observations show an
asymmetric distribution of flux ratios, with a long tail towards high
radio luminosities where only a small fraction of quasars
lie. However, firm proof on whether this is due to a bimodal
distribution or simply an asymmetric continuous radio luminosity
distribution, is still elusive, as attempts to hunt for a bimodality
in the radio loudness distribution of quasars have provided
contradictory conclusions. Some authors have argued that the
bimodality exists \citep[e.g.][]{Ivezic2002,White2007} while others
claim there is no bimodality
\citep[e.g.][]{Cirasuolo2003b,Cirasuolo2003a,Balokovic2012}. An
inherent problem is the selection biases that exist in these studies
due to the definition of radio-loudness using optical
and radio information, and the use of flux-limited samples at both
radio and optical wavelengths. Furthermore, the use of different
observing bands can give varying results \citep{Ivezic2002}. Lacking
the answer to such a fundamental question about the nature of RL and
RQ quasars means we cannot obtain a complete understanding of the
physical mechanisms at play. Moreover, this gap in our knowledge has
significant consequences for theories of galaxy formation and
evolution given the strong evidence for a close relationship between
the growth of the central SMBH and the evolution of its surrounding
host galaxy \citep[e.g.\ see reviews by][]{Fabian2012,HeckmanBest},
such as the correlations observed between the mass of the central SMBH
and properties of the host galaxy's bulge
\citep[e.g.][]{Ferrarese2000,Gebhardt,Marconi2003,Haring2004}.

The source of the radio emission from `RQ' quasars is also a debated
issue \citep[see][for a review]{Panessa2019}. Star-formation (SF)
produces free-free emission from HII regions as well as synchrotron
radiation from electrons accelerated to relativistic speeds in
supernova remnants \citep{Condon1992}. As the hosts of quasars are
often star-forming galaxies \citep[see review by][]{HeckmanBest}, the
question then becomes whether SF in the host galaxy is sufficient to
account for the observed radio emission from RQ quasars. Some studies
have found that the radio emission from some RQ quasars could be
explained solely by SF \citep[e.g.][]{Kimball2011,Condon_2013}. Others
suggest that SF is not sufficient and, hence, that the majority of the
radio emission from RQ quasars must come from the AGN
\citep[e.g.][]{Zakamska2016,White2015,White2017}, either in the form
of small-scale jets \citep{Cirasuolo2003b}, or from other processes
that produce weak radio emission such as AGN-driven winds or disk
coronal activity \citep[e.g.][]{Laor2008}. In many cases, high
resolution radio maps of RQ quasars have detected non-thermal radio
core emission or emission coming from extended jet-like structures
\citep[e.g.][]{Kukula1998,Blundell1998,Leipski2006,Klockner2009,Maini2016,Ruiz2016,Jarvis2019,Hartley2019}.
It could be the case therefore that jets are a feature in all quasars
but are often unresolved. This would not be surprising since such low
luminosity radio jets are commonly seen in massive galaxies
\citep[although these radio-AGN are not optically classified as
  quasars; e.g.][]{HeckmanBest,Mingo2019,Baldi2021}, with
\citet{Sabater2019} finding evidence that locally they are essentially
always present in the most massive galaxies. \citet{Mancuso2017}
attempted to explain the abundances of SF-dominated and jet-dominated
RQ quasars using a model of in-situ evolution whereby the origin of
the dominant radio emission changes as the black hole grows.

Many studies have investigated the SF component of quasars, and in
particular how the star-formation rate (SFR) relates to the optical or
X-ray luminosity of the quasar, and how it evolves with redshift. The
relation between SFR and quasar luminosity is particularly interesting
because it effectively relates the growth rate of the SMBH to that of
the galaxy around it, and hence the build-up of the black hole mass
versus bulge mass relation. \citet{Netzer2009} and
\citet{Bonfield2011} both suggested a strong correlation between AGN
luminosity and star-formation rate using AGN from the Sloan Digital
Sky Survey \citep[SDSS;][]{York2000}. More recent studies of (often
lower luminosity) quasars at higher redshifts find that the SFR
increases strongly with increasing redshift out to at least $z\sim2$,
but that any trend of SFR with quasar luminosity is either weak
\citep[e.g.][]{Harrison2012,Azadi2015,Stanley2017,Stemo2019} or
insignificant \citep[e.g.][]{Rosario2012,Mullaney2012,Stanley2015}
once the redshift effects are accounted for. However, the quantitative
details remain widely debated. Furthermore, there is an exception to
this result at the highest quasar luminosities, where most studies
find that SFR does correlate with AGN luminosity
\citep[e.g.][]{Shao2010,Rosario2012,Harris2016,Lanzuisi2017}; this has
been attributed to the triggering of these quasars by major mergers of
galaxies.

The physical mechanisms that produce radio jets in RL (and at least
some RQ) quasars are also still uncertain. Some proposed theoretical
models are compelling but lack observational
confirmation. \citet{Blandford1977} proposed that a spinning BH,
threaded by magnetic field lines, can produce anti-parallel jets of
energy, while \citet{Blandford1982} proposed that outflows of matter
and energy from a rotating disk of gas accreting on to a spinning
black hole can form jets when certain conditions regarding the
orientation of the magnetic field with respect to the disk are
met. BHs can get spun-up by recent galaxy mergers and perhaps by
accretion events \citep[e.g.][]{Dotti2013}, and \citet{Martinez2011}
found that using this assumption they were able to explain the radio
luminosity functions of both high- and low-excitation radio
galaxies. However, no conclusive evidence has yet been found that
suggests the BH spin is involved in generating radio jets in quasars,
due to the extreme difficulty in measuring the spin of the BH
\citep[e.g.\ see the contradictory results for galactic-scale black
  holes from][]{Steiner2013,Russell2013}. \citet{Wilson1995} argued
that BH spin must provide the fundamental distinction between RL and
RQ quasars given that they found no dependence of radio loudness on
other physical parameters, such as the mass of the BH and the
accretion rate. However, while some later studies back up these
findings \citep[e.g.][]{Woo2002}, many others do find strong
dependencies of radio loudness on the BH mass
\citep[e.g.][]{Laor2000,Lacy2001,Dunlop2003,McLure2004,Best2005}. RL
quasars are also found to be more highly clustered than RQ quasars
\citep{Retana2017}, suggesting that larger-scale environment may play
a role.

\citet{vanVelzen2013} find a very tight relationship between jet and
disk luminosities in RL quasars, and infer that if BH spin is a major
factor in the jet power then the RL quasars must all have very similar
spin values, with then a wide gap in spin to the RQ population; this
would require a strong dichotomy in the quasar
population. \citet{Sikora2013b} argue that variations in the strength
of the magnetic flux threading a spinning black hole may instead be
the primary factor that controls the strength of radio jets, while
several recent works \citep{Klindt2019,Rosario2020,Fawcett2020} have
found that red quasars have a factor $\sim 3$ higher radio-detection
fraction than blue quasars \citep[see
  also][]{Richards2003,White2007,Calistro2020}, and argue that an
evolutionary sequence may be occurring. Similarly,
\citet{Morabito2019} found a higher radio-loud fraction in broad
absorption line quasars and evidence for a link between the radio
activity and an outflow phase.

Hence, there are several prominent questions still at large:
\begin{enumerate}
    \item Are all quasars part of the same population?
    \item Is SF sufficient to account for the radio emission observed
      from RQ quasars or are small-scale radio jets prevalent?
    \item How does the SF rate depend on quasar properties (redshift,
      luminosity) and what does this tell us about the triggering of
      the quasar activity?
    \item What are the physical mechanisms that influence the prevalence 
      or strength of radio jets in quasars?
\end{enumerate}

The LOw Frequency ARray \citep[LOFAR;][]{vanHaarlem} Two-metre Sky
Survey \citep[LoTSS;][]{Shimwell2017} is an ongoing radio survey of the
northern sky in the frequency range 120-168 MHz, detecting an order of
magnitude higher sky density of sources than any previous large-area
radio survey (see Section~\ref{Sec:radio} for more details). LoTSS is
providing deep radio imaging of large samples of quasars. Furthermore,
in the low-frequency radio regime, any extended radio structures
(which are likely to have steeper radio spectral indices) are more
prominent than at higher frequency, and therefore potential biasing
effects due to Doppler boosting are much less pronounced than at GHz
frequencies.

Recently, \citet{Gurkan2019} used the LoTSS data release 1
\citep[DR1;][]{LoTSS_DR1, Williams, Duncan2019} data over the
Hobby-Eberly Telescope Dark Energy Experiment \citep[HETDEX;
][]{Hill2008} Spring Field region and the LOFAR Herschel-ATLAS North
Galactic Pole survey \citep[H-ATLAS NGP;][]{Hardcastle2016} to examine
the low-frequency radio properties of optically selected quasars from
the SDSS Baryon Oscillation Spectroscopic Survey
\citep[BOSS;][]{Dawson2013}. The wide area and high sensitivity of the
LoTSS DR1 allowed \citet{Gurkan2019} to determine radio luminosities,
or place meaningful upper limits on these, for tens of thousands of
quasars. They investigated how the radio loudness of these quasars
depended on other galaxy and BH parameters such as BH mass, optical
luminosity, radio luminosity, redshift, and Eddington
ratio. Given their results, \citeauthor{Gurkan2019} favour the
scenario where AGN jets (of a wide range of powers) and star
formation-related processes both contribute to the radio emission
observed from quasars and that there is no RL/RQ dichotomy, but rather
a smooth transition between the regimes where each of the two
processes dominate.

We aim to build upon the results of \citeauthor{Gurkan2019} by taking
the simple approach of constructing and testing a numerical,
two-component model of the radio flux densities of quasars. The model
is a superposition of the two expected sources of radio emission from
galaxies, the AGN (jets) and the SF, each modelled from physical
prescriptions. The model implicitly assumes that no intrinsic
bimodality exists but rather there is a smooth transition from a
star-formation dominated to a jet-dominated regime as the radio jet
power increases. Such an approach allows us to generate simulated
samples through Monte Carlo realisations that can be compared to
observed data. Quantifying the validity of the model and constraining
its parameters will provide information relevant to questions (i) and
(ii) above, while investigating how the model parameters change as a
function of properties of the quasar such as redshift, optical
luminosity (for which we use the absolute \textit{i}-band magnitude
as a proxy) and BH mass provides input into questions (iii) and (iv).

The outline of the paper is as follows.  Section~\ref{Sec:data}
outlines the data used during this research to build a large sample of
quasars used to validate and constrain the model. The two-component
model of the radio luminosity distribution of quasars is detailed in
Section~\ref{Sec:model}.  Section~\ref{Sec:KSresults} presents the
results found, and the physical interpretations of these results are
discussed in Section~\ref{Sec:discussion}. Finally, a summary of our
conclusions is given in Section~\ref{Sec:conclusion}.  Cosmological
parameters are taken to be $(\Omega_m,\Omega_{\Lambda}) = (0.3,0.7)$
and $H_0 = 70$ km s$^{-1}$ Mpc$^{-1}$.

\section{Data} \label{Sec:data}

\begin{comment}
In this section, we describe the data used during this
research. Radio data from the LOFAR Two-metre Sky Survey (LoTSS)
(Section~\ref{Sec:radio}) are used to analyse the radio properties of
quasars observed in the optical from the Sloan Digital Sky Survey
(SDSS; Section~\ref{Sec:SDSS}) across a range of redshifts:
$0<z<6$. Section~\ref{Sec:radio} and \ref{Sec:SDSS} briefly detail the
two surveys respectively, while Section~\ref{Sec:sample} discusses our
sample of cross-matched quasars. Finally,
Section~\ref{Sec:prop_quasars} describes how we yield useful
properties of the quasars, namely: radio flux densities, accretion
rates and virial BH mass estimates.
\end{comment}

\subsection{LOFAR Two-metre Sky Survey (LoTSS)}
\label{Sec:radio}

LoTSS\footnote{lofar-surveys.org} \citep{Shimwell2017} is an ongoing
radio survey with a frequency range of 120--168 MHz\footnote{The
  central frequency of LoTSS band is 144 MHz, but the
  sensitivity-weighted mean frequency varies with position due to the
  frequency-dependent primary beam size. For simplicity, we use 150
  MHz throughout the paper to refer to the LoTSS frequency.}, target
rms sensitivity of $\sigma_{\rm 150MHz} < 100\mu$Jy\,beam$^{-1}$,
image resolution of $6^{\prime \prime}$ and positional accuracy for
brighter sources of better than $0.2^{\prime \prime}$. Once completed,
LoTSS will have surveyed the entire northern sky but here we make use
of LoTSS data release 1 \citep[DR1;][]{LoTSS_DR1}, which covers 424
sq.\ deg.\ in the region of the HETDEX Spring Field (RA: 10h45m00s -
15h30m00s, DEC: $45^{\circ}00^{\prime}00^{\prime \prime}$ -
$57^{\circ}00^{\prime}00^{\prime \prime}$) to a median 150-MHz rms
sensitivity of 71$\mu$Jy\,beam$^{-1}$.

LoTSS DR1 includes the radio images of the relevant region as well as
a source catalogue \citep{LoTSS_DR1}, where the Python Blob Detector
and Source Finder \citep[PyBDSF;][]{PyBDSF} algorithm has been
used to catalogue sources. However, the PyBDSF catalogue produced
will contain a number of sources that were associated incorrectly. The
reasons for this could be: the blending of physically distinct sources
into a single catalogue entry; the separation of the components of
extended sources into different PyBDSF catalogue entries; spurious
emission or artefacts. To account for this, \citet{Williams} present a
value-added catalogue in DR1 where significant effort (both
statistical techniques and extensive visual analysis known as LOFAR
Galaxy Zoo) has gone into ensuring, as much as possible, that the
catalogue is a true representation of the radio sources in the
relevant region. Furthermore, \citeauthor{Williams} provided
optical/infrared counterpart identifications (where detected) for all
of the LoTSS sources, making use of optical and infrared data from the
Panoramic Survey Telescope and Rapid Response System
\citep[Pan-STARRS][]{PanSTARRS} $3\pi$ survey and the Wide-field
Infrared Survey Explorer \citep[WISE;][]{WISE}. The detailed processes
implemented to produce such a catalogue can be found in
\citet{Williams}.

\subsection{Sloan Digital Sky Survey quasar sample}
\label{Sec:SDSS}

Optical data for quasars from the fourteenth data release
\citep[DR14Q;][]{Myers2015,SDSS} of the SDSS were obtained, using the
catalogue detailed in \citet{SDSS}. The catalogue of over half a
million quasars is described as a \textquote{superset} that is a
compilation of all spectroscopically confirmed quasars from SDSS-I,
II, III and IV \citep[see also][for earlier selection
  criteria]{Richards2002,Ross2012}.  It was also necessary however, to
obtain the SDSS-DR7 quasar catalogue (DR7Q) that consists of all
spectroscopically confirmed quasars from SDSS-I/II
\citep{Schneider2010} in order to obtain the target selection flags of
the 79,487 quasars that were not re-observed as part of SDSS-IV. The
target selection flags allowed us to identify and remove quasars that
were selected for observation solely based on their radio emission, in
order to mitigate any selection bias (see Section~\ref{Sec:sample}).

Absolute \textit{i}-band magnitudes were also obtained from the DR14Q
catalogue. We convert the magnitude given in the catalogue (a
magnitude \textit{K}-corrected to $z=2$), to a magnitude
\textit{K}-corrected to $z=0$, $M_i(z=0)$, assuming an optical
spectral index of 0.5 \citep{Richards2006}. Galactic extinction
corrections for the \textit{i}-band were also applied. The extinction
correction for each quasar, obtained from the \citet{Schlafly} dust
maps, was given in the DR14Q catalogue.

\subsection{Our Sample} \label{Sec:sample}

Our aim was to build a sample of quasars observed in the optical by
the SDSS, which provides information on their redshift and optical
luminosity, and to combine this with the radio properties of the
quasars, such as their integrated flux density, from LoTSS. We started
with the $\sim51,000$ quasars within the LoTSS DR1 RA and Dec limits
from the SDSS DR14Q catalogue described in Section~\ref{Sec:SDSS}. The
following cuts were then applied in order to ensure a robust and
unbiased analysis:

\begin{enumerate}
    \item Sources brighter than $M_i = -40$ were removed. These were
      likely misidentified during the automated process that generated
      the DR14Q catalogue.
    \item The sample was limited out to a redshift of $z=6$, as beyond
      that the contamination from unreliable redshifts is high (in
      practice our analysis was restricted to even lower redshifts by
      sample size limitations).
    \item Sources that lay outside the LoTSS coverage, or
      fell within gaps of the LoTSS mosaics, were removed, due to the
      lack of radio data.
    \item Finally, a further 199 sources that had target selection
      flags in SDSS DR7Q or DR14Q that indicated they were included in
      the spectroscopic sample {\it solely} because of their radio
      emission were removed, to mitigate any possible bias (otherwise,
      in regions outside of the SDSS colour-space selections,
      radio-bright quasars would be preferentially included,
      potentially biasing the results).
\end{enumerate}

\begin{figure}
    \centering
    \includegraphics[width=\columnwidth]{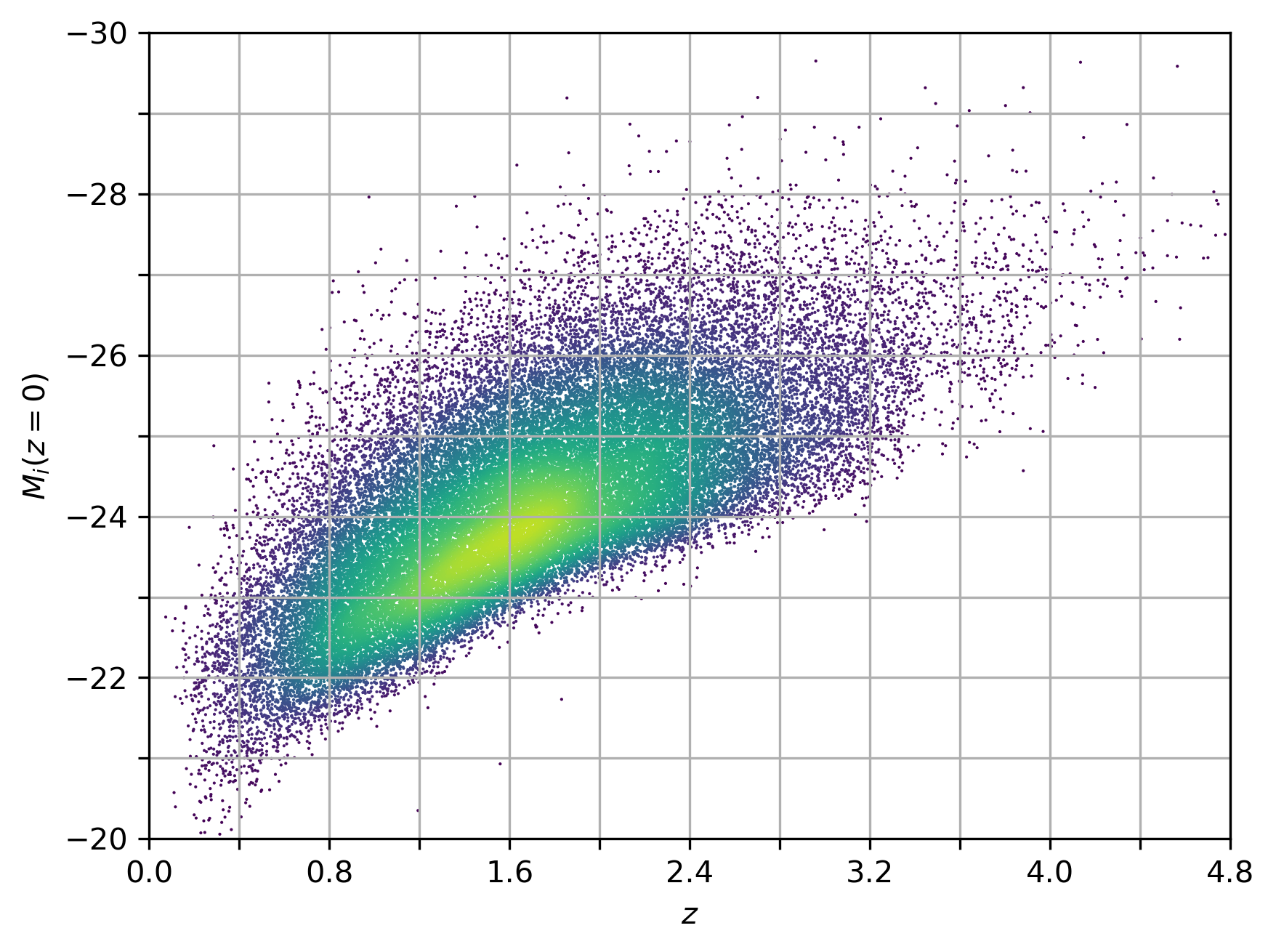}
    \caption{The distribution of the final sample of quasars in $M_i -
      z$ space. The colour scheme characterises the density (Gaussian
      kernel density estimation) of sources in $M_i - z$ space. A grid
      is included with grid lines in steps of $\Delta z = 0.4, \Delta
      M_i = 1$ that indicates where subsamples of the main sample were
      produced for comparison with our model (see Section~\ref{Sec:model}).}
    \label{fig:L-z_plot}
\end{figure}

The final sample had a size of 42,601 quasars. The distribution of
these quasars in $M_i - z$ space is shown in Fig.~\ref{fig:L-z_plot}.
As expected, the sample probes systematically higher optical
luminosities at higher redshifts, but still has a good dynamic range
in optical luminosity at each redshift. 

\subsection{Properties of the Quasars} \label{Sec:prop_quasars}

\subsubsection{Radio Flux Densities} \label{Sec:cross-match}

To determine radio flux densities for our quasar sample, we first
cross-referenced the coordinates of the two survey catalogues, LoTSS
and SDSS. It should be noted here that the coordinates of the optical
identification given in the LoTSS value-added catalogue (i.e.\ that of
the cross-matched Pan-STARRS or WISE host galaxy) were used rather
than the less accurate radio-derived coordinates from PyBDSF; the
Pan-STARRS coordinates are aligned to the SDSS co-ordinates to
typically much better than an arcsecond.

\begin{figure}
    \centering
    \includegraphics[width=\columnwidth]{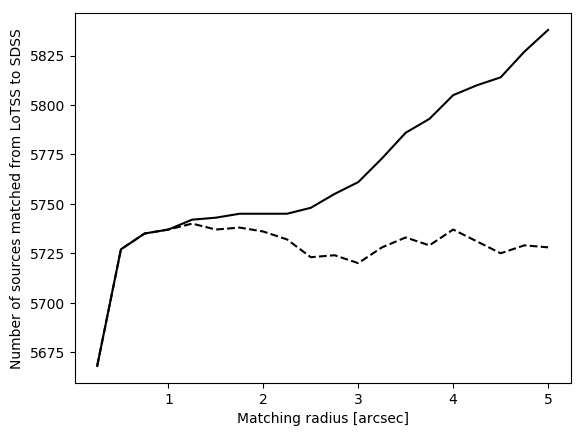}
    \caption{The solid line gives the raw number of sources matched
      and the dashed line gives the number after statistically
      correcting for the expected random contamination. The plateau in
      the corrected matched counts begins at around $1.5^{\prime
        \prime}$.}
    \label{fig:match}
\end{figure}

Fig.~\ref{fig:match} shows the number of quasars matched to a radio
source (their nearest) as a function of the maximum cross-matching
angular distance. To estimate the contamination one would observe for
a given matching radius, we select $N$ random locations within the
LoTSS DR1 coverage and measure the number of random matches as a
function of radius. Fig.~\ref{fig:match} also shows the number of
matches out to each cross-matching radius after correcting for this
random contamination. From this, the maximum cross-matching radius was
chosen to be $1.5^{\prime \prime}$. This resulted in just under 5750
direct matches with a predicted contamination of $\sim0.1\%$. For
these cross-matched sources, we extract integrated flux densities (and
associated uncertainties) directly from the \citet{Williams}
value-added catalogues. These integrated flux densities correctly
incorporate any extended radio structures due to the source
association process.

For the remaining optical quasars that were undetected in the LoTSS
catalogue, a flux density was extracted directly from the LoTSS
mosaics, using the image value in the mosaics at the coordinates of
the quasars given in DR14Q. The uncertainty on this was extracted at
the same location from the LoTSS rms noise maps. Although these
sources are individually below the 5$\sigma$ catalogue S/N limit and
so have measured flux densities dominated by the noise (with many
having negative values), it is nonetheless expected that the genuine
low significance emission will lead to considerable information being
present in the exact flux density distribution \citep[cf.][see also
  Fig.~\ref{fig:flux_example}]{Roseboom2014,Malefahlo2020}. An
underlying assumption of extracting flux densities directly from the
radio maps is that these faint sources must be compact compared to the
$6^{\prime\prime}$ LoTSS beam (such that the peak flux density traces
the integrated flux density); this is a reasonable approximation as
any radio emission from these faint sources will primarily be
star-formation on galaxy scales, radio cores and/or small-scale
jets. It should be noted that this faint emission will not have been
properly cleaned in the radio imaging step, which may lead to a slight
under-estimate of the true flux densities.

For a small number (a few tens) of quasars, despite the lack of a
LoTSS catalogue match, the radio flux density extracted from the
mosaics had a flux density with greater than 5$\sigma$ significance. A
selection of these sources were examined visually, and found to
represent a mixture of cases such as sources PyBDSF had failed to
detect, incorrect LoTSS IDs and quasars overlying the extended radio
emission of other sources. In the former cases, the integrated flux
density may be underestimated if the radio source is extended, while
in the latter case it will be overestimated. As there were relatively
few of these sources, and flux densities may be biased in either
direction, no attempt was made to correct these errors.

\subsubsection{Bolometric Luminosity and Accretion Rates}
\label{Sec:accretion}

It is also useful at times during our analysis to relate the absolute
\textit{i}-band magnitude\footnote{Note that the $M_i$ selection could
  lead to an under-estimate of true luminosities, and hence accretion
  rate, for reddened quasars. However, the fraction of quasars that
  are heavily reddened in the $i$-band is small and so the effect is
  not expected to be significant.} to a bolometric luminosity ($L_{\rm
  bol}$) and to an estimate of the growth rate of the SMBH,
$\dot{M}_{\rm BH}$. To do this, we first relate the absolute
\textit{i}-band magnitude to the absolute \textit{b}-band magnitude
using the relation determined empirically from a sample of 1046
quasars by \citet{Richards2006}: $M_{b_J} - M_i(z=2) =
0.66\pm0.31$. This allows us to estimate the bolometric luminosity
applying the empirically-derived relation given in \citet{McLure2004}:

\begin{equation}
    M_B = -2.66(\pm0.05)\ \mathrm{log}[L_{\rm bol}/\mathrm{W}] + 79.36(\pm1.98).
\end{equation}

\noindent Although there may be slight discrepancies between the
\textit{b}-band filters used to calibrate the above relations, the
relations should give a sufficiently good estimate. The bolometric
luminosities of the quasars in the sample range from about $10^{38}$
to $10^{40}$\,W.

The BH growth rate is then given by

\begin{equation}
  \dot{M}_{\rm BH} = \frac{(1 - \epsilon) L_{\rm bol}}{\epsilon  c^2}
\end{equation}
 
\noindent where $\epsilon$ is the efficiency at which the rest-mass of
the accreting material is converted to energy; we assume a typical
value of $\epsilon=0.1$.

\subsubsection{Virial Black Hole Masses} \label{Sec:BHmass}

To further our analysis of the radio loudness distribution of the
sample beyond dependencies on optical luminosity and redshift, we have
also obtained virial BH mass estimates. These have been obtained from
catalogues provided by \citet{Shen2011} and \citet{Kozlowski2017}.
These catalogues only include quasars up to the SDSS data release 12,
and so do not include all of the quasars in our sample, but should
represent a relatively unbiased subsample. The methods used to obtain
the black hole mass estimates are briefly discussed here.

\citet{Shen2011} present properties of quasars in the SDSS DR7Q
catalogue based on spectral fits. We use the measurements of the
virial BH mass based on the broad MgII and CIV emission lines, which
have been calibrated as virial BH mass estimators. Several estimates
are given for varying calibrations of the MgII line by
\citet{Shen2011}. \citet{Kozlowski2017} make use of the broadband,
extinction-corrected magnitudes from quasars in the SDSS DR12Q
catalogue to derive monochromatic luminosities that can then be
combined with the broad emission line widths of MgII and CIV to
estimate BH masses. 

The \citet{Shen2011} and \citet{Kozlowski2017} catalogues were matched
to the SDSS-LoTSS sample described in Section~\ref{Sec:sample}, using
positional cross-match (all matches being found within
$\sim0.6^{\prime \prime}$, with essentially no contamination). The two
sets of black hole mass estimates are broadly comparable, and the
precise choice of which to use has no qualitative effect on the
results that we obtain. Where available, we use the estimates provided
by \citet{Shen2011} and the weighted-mean of the available estimates
was taken. Otherwise, the estimates provided by \citet{Kozlowski2017}
were used, where the MgII estimate was prioritised over the CIV
estimate because \citet{Kozlowski2017} uncovers a bias with
measurements from the CIV line. Of the 42,601 sources in the
SDSS-LoTSS sample, 24,096 have estimates for the virial BH mass. The
derived black hole masses typically range from $10^{8.5}$ to $10^{9.5}
{\rm M}_{\odot}$ (although at lower redshifts the lower-luminosity
quasars have lower black hole masses, down to $10^{7.5} {\rm
  M}_{\odot}$).

\section{A Two-component Model for Radio Emission}
\label{Sec:model}

We started from the simple approach of building a two-component model
of the radio luminosity distribution of quasars. The model assumes
that two sources (AGN and SF) contribute to the observed radio
emission of every quasar. Thus, in our model, all quasars are assumed
to display star-formation activity at some level \citep[as would be
  expected, given the large gas content that is present in order to
  fuel the black hole accretion, and in line with many studies of
  quasar host galaxies; e.g.][]{Shao2010,Floyd2013,Harris2016}. In
Section~\ref{Sec:SFcomp} we describe how we assign an SFR to each
quasar, drawing from an inferred Gaussian distribution.  In addition,
in our model all quasars are assumed to possess radio jets. As we
motivate in Section~\ref{Sec:AGNComp}, the jet luminosity is allowed
to vary in strength from the very powerful radio jets seen in the most
radio-loud quasars down to the very weak small-scale radio jets that
have been observed in high angular resolution, sensitive radio images
of some radio-quiet quasars. The large range of possible
jet powers is the primary factor that sets the overall radio
luminosity of the system, and determines whether the AGN or the SF is
the dominant source of radio emission.

We create simulated samples of quasars from our model through Monte
Carlo realisations, summing the radio luminosity contributions from
the SF and AGN components, converting these to a radio flux density,
and adding noise (Section~\ref{Sec:TotalFlux}). We then compare these
with the observed distribution of quasar flux densities in order to
determine the validity of our model prescription for the sources of
radio emission and to constrain the model parameters (as outlined in
Section~\ref{Sec:ks}).

As it is expected that the strength of both the star formation and the
jet component may be dependent on the optical luminosity of the
quasar, and may vary with redshift, we carry out this comparison on
subsamples of quasars produced by separating the main sample in $M_i -
z$ space by the grid lines shown in Fig.~\ref{fig:L-z_plot}. We
separately analyse each subsample in each grid square that hosts over
500 quasars. Analysing the distribution in the 2-dimensional $L_{\rm
  opt} - z$ space is preferred to marginalising the distributions, as
\citet{Jiang2007} showed that the strong correlation between $L_{\rm
  opt}$ and $z$ can lead to marginalised studies obtaining inaccurate
results. Within each such grid square, the optical luminosity and
redshift are reasonably constant ($\Delta M_i = 1; \Delta z = 0.4$),
so fitting the same model parameters for the star formation and jet
components for all quasars within each bin should produce robust
results. By analysing the results from the different individual grid
squares separately, we are able to recover detailed information as to
how our model parameters for the SF and jet components vary as a
function of cosmic time and optical luminosity.

\subsection{SF Component}
\label{Sec:SFcomp}

The radio luminosity function of star-forming galaxies is
  often modelled as a broken power-law, where the wide distribution of
  star-formation rates arises from the large range of stellar masses
  of the star-forming galaxies, combined with the tight relation
  (scatter $\sim$0.2-0.35 dex) that star-forming galaxies show between
  their SFRs and their stellar masses \citep[often called the
    star-forming main sequence; e.g.][]{Noeske2007,Elbaz2007}. Low
  star formation rates also arise from the quiescent galaxy
  population.

The host galaxies of powerful AGN are typically both massive
\citep[e.g.][]{McLure1999,Best2005} and star-forming
\citep[e.g.][]{Kauffmann2003}. Investigations have found that the host
galaxies of radiatively-efficient (quasar-like) AGN mostly lie on or
above the SFR-mass relation \citep[e.g.][]{Mainieri2011,HeckmanBest},
avoiding the quiescent galaxy population\footnote{This is different
  for the radiatively-inefficient, jet-mode AGN which are mostly
  located in massive quiescent galaxies \citep[e.g. see review
    by][]{HeckmanBest}, but by definition these jet-mode AGN do not
  host quasars.}. Since they are all high-mass star-forming galaxies,
the SFRs of quasar host galaxies would therefore be expected to have a
much narrower distribution than a full power law.  Based on this, we
model the radio emission of the SF component from each quasar (in a
given bin of redshift and optical luminosity) as being drawn from a
Gaussian distribution (in log space) with two free parameters: the
mean, $\mathrm{log}(L_{\mu}/[\mathrm{W\ Hz^{-1}}])$ at the frequency
of LOFAR ($\sim$150\,MHz), and the standard deviation,
$\sigma/\mathrm{dex}$.

The normalisation of the star-forming main sequence is known to evolve
strongly with redshift due to the higher availability of gas in the
early Universe \citep[e.g.][]{Speagle2014}: the mean star
formation rates of massive galaxies increases by a factor $\sim 30$
from $z=0$ to $z \sim 2$, in line with the evolution of the cosmic
star formation rate density \citep[e.g. see review
  by][]{Madau2014}. Furthermore, as discussed earlier, many studies
have investigated how the typical SFR of quasars varies with their AGN
luminosity, finding different results. For this reason, the two
Gaussian parameters (mean and standard deviation) are allowed to take
different values in different bins of redshift and luminosity. In each
Monte Carlo realisation, a random luminosity for the SF component of
each quasar ($L_{\rm SF}$) is drawn from this Gaussian
distribution. The determination of the best-fitting values for
$L_{\mu}$ and $\sigma$ in the different redshift and optical
luminosity bins then reveals how the SF component evolves across
cosmic time and how it connects to the black hole accretion rate.

To yield physical information about the system, it is useful to relate
the mean luminosity of the SF component to a SFR, $\Psi$. Such
calibrations at 150 MHz have been provided recently by
\citet{Brown2017} based on data from the alternative data release of
the TFIR GMRT Sky Survey \citep[TGSS;][]{Intema2017} and by
\citet{Calistro2017} and \citet{Gurkan2018} based on LoTSS data. The
Brown and G\"urkan relations (calibrated with a Chabrier 2003 initial
mass function) \nocite{Chabrier2003} are
\footnote{Note that there is an error in the conversion from
  $L_{H\alpha}$ to SFR for a Chabrier IMF in footnote (b) of Table 3
  of \citet{Brown2017}, where the Salpeter to Chabrier IMF conversion
  factor appears to have been inversely applied; the values provided
  here use the correct conversion.}

\begin{equation}
  \log \left(\frac{\Psi}{{\rm M}_{\odot} \rm{yr}^{-1}}\right) = 0.86
  \left[\log \left(\frac{L_{150}}{\rm{W}\,\rm{Hz}^{-1}} \right) - 21.97 \right]
\end{equation}
  
\begin{equation}
  \log \left(\frac{\Psi}{{\rm M}_{\odot} \rm{yr}^{-1}}\right) = 0.93
  \left[\log \left(\frac{L_{150}}{\rm{W}\,\rm{Hz}^{-1}} \right) - 22.06 \right]
\label{eqn:sfr_l150}
\end{equation}

\noindent which agree to within 0.1 dex at the typical luminosities of
the quasars in our sample ($L_{150} \sim 10^{23} -
10^{24}$W\,Hz$^{-1}$). Here we use the latter relation, which has been
calibrated using LOFAR data. To relate the width of the Gaussian,
which is in $\log L$ space, to a width in $\log \Psi$ space,
Eqn.~\ref{eqn:sfr_l150} gives $\sigma_{\Psi} = 0.93 \sigma$.

\subsection{AGN Component}
\label{Sec:AGNComp}

The radio luminosity function of `radio-loud' quasars (i.e.\ those with
a dominant AGN component) is often modelled as a broken power-law
\citep[e.g.][]{DP90,Kaiser2007,Kimball2011,Best2014}. However, the
radio luminosity function maps out the full source population, and it
is likely that quasars of different optical luminosity may dominate
different parts of this distribution. Furthermore, a broken power-law
model requires two additional free parameters over a single power-law
formalism, and the data do not justify these additional parameters:
when attempting to model the AGN (jet) component in given redshift and
radio luminosity bins as a broken power law, we found that we could
not constrain the slope above the break luminosity and there were
strong degeneracies between other parameters. Instead, a single
power-law formalism was found to be sufficient to describe the AGN
component of the model. 

To model the AGN component, we therefore draw a luminosity randomly
from a single power-law distribution, with probability distribution
function (PDF)

\begin{equation}
    \rho(L) = \rho_0 L^{-\gamma},
\end{equation}

\noindent where $\rho_0$ is the normalisation and $\gamma$ is the
slope. The PDF was defined such that $\rho$ has units $(\Delta
L)^{-1}$. Defining $\rho$ as such means that the integral of the PDF
must adhere to:

\begin{equation}
    \int^{L_u}_{L_l} \rho(L)dL = \int^{L_u}_{L_l} \rho_0 L^{-\gamma}dL = 1
    \label{eq:pdf_int}
\end{equation}

\noindent where $L_u$ is the maximum radio luminosity obtained by any
radio quasar, and $L_l$ is the minimum jet luminosity of the
quasars. The choice of $L_u$ is not critical, as the integral of the
model in the range $L_u \rightarrow \infty$ will be negligible
provided that $L_u$ is sufficiently large. In practice, we set
$\mathrm{log}(L_u/[\mathrm{W\ Hz^{-1}}]) = 30$, above which we do not
expect to see any sources. The lower limit, $L_l$, is then fixed, for
a given normalisation and slope, by Equation~\ref{eq:pdf_int}.

To set the normalisation of the power-law, $\rho_0$, in to more
physically intuitive units, we define the quantity $f$ as the fraction
of the integral of the PDF at luminosities brighter than $L_f$, where
$L_f$ was set as $\mathrm{log}(L_f/[\mathrm{W\ Hz^{-1}}])= 26$
(choosing a different value just leads to a scaling of $f$, but
does not affect the trends seen). The value of $L_f$ was chosen to be
high enough that jet emission will dominate and SF will be negligible,
and thus variation of $f$ maps directly onto the variation in the
fraction of high-power jet-dominated radio sources. The two free
parameters in the model are now $f$ and $\gamma$.

It can be shown that,

\begin{equation}
    \rho_0 = \frac{(1-\gamma)f}{L_u^{1-\gamma} - L_f^{1-\gamma}}.
\end{equation}

\noindent Asserting equation~(\ref{eq:pdf_int}) now means $L_l$ can be
computed:

\begin{equation}
    L_l = [(1-\frac{1}{f})L_u^{1-\gamma} +
      \frac{1}{f}L_f^{1-\gamma}]^{\frac{1}{1-\gamma}}.
\end{equation}

\noindent For the best-fit parameters of the model, we typically
observe lower limits in the range
$\mathrm{log}(L_l/[\mathrm{W\ Hz^{-1}}]) \approx 20-22$.  By
comparison, \citet{Mauch2007} compare their integrated 1.4\,GHz radio
luminosity function with the space density of massive galaxies and
conclude that the radio luminosity function must turn down below about
$L_{\rm 1.4GHz} \approx 10^{19.5}$W/Hz, and \citet{Cattaneo2009} come
to a similar value ($10^{19.2}$W/Hz) by comparison with the space
density of massive black holes. These values (which correspond to
about $10^{20}$W/Hz at 150\,MHz assuming a typical spectral index,
$\alpha \sim 0.7$) are dervied assuming that the same
limiting radio luminosity holds for all galaxies. \citet{Sabater2019}
investigate AGN fractions as a function of stellar mass, and find that
the cumulative AGN fractions reach 100 per cent (and hence the
luminosity function must turn over) at around $L_{\rm 150MHz} \sim
10^{22}$W/Hz at the highest stellar masses, with the cut-off
luminosity decreasing with decreasing stellar mass.  The range of
lower limits determined for the model, $L_l \sim 10^{20} -
10^{22}$W/Hz, agrees well with these observations, giving confidence
that the model is producing sensible results.

To randomly sample from the single power-law, the inverse cumulative
method was implemented. Briefly, this method involves building the
cumulative distribution function (CDF):

\begin{equation}
    \rm{CDF}(L) = \int^{L}_{L_l} \rho(L)dL \equiv Y
\end{equation}

\noindent where $Y$ is a uniformly distributed variate on (0,1), and
then inverting the function to solve for $L$. The luminosity variates
of the AGN component can therefore be sampled from:

\begin{equation}
    L \equiv L_{\rm AGN} = [(L_u^{1-\gamma} - L_l^{1-\gamma})Y +
      L_l^{1-\gamma}]^{\frac{1}{1-\gamma}}.
\end{equation}

\subsection{Total Simulated Flux Density}\label{Sec:TotalFlux}

For each quasar, for a given set of values of the four free parameters
in the model ($L_{\mu}$, $\sigma$, $f$ and $\gamma$) a luminosity is
randomly drawn (independently) for each of the components (i.e.\ a SF
luminosity and an AGN luminosity). The two luminosities are then added
together to give a total luminosity:

\begin{equation}
    L_{\rm model} = L_{\rm AGN} + L_{\rm SF}.
\end{equation}

\noindent Using the known redshift of the quasar, a simulated flux
density can then be computed. Remembering that we need to apply a
\textit{K}-correction, assuming $L\propto \nu^{-\alpha}$, the flux
density can be computed using:

\begin{equation}
    S(\nu_0) = \frac{L(\nu_0) (1+z)^{1-\alpha}}{4\pi D_L^2},
    \label{eq:simflux}
\end{equation}

\noindent where $\nu_0$ is the frequency of LOFAR observations,
$\alpha§=0.7$ is the radio spectral index \citep[e.g.][]{Calistro2017}
and $D_L$ is the luminosity distance.  Gaussian noise is then added to
simulate the observation, where the width of the Gaussian is taken to
be the rms extracted from the LoTSS mosaic at the position of that
quasar. This process is repeated for all quasars to produce a single
representation of the simulated flux density distribution for the
quasar population. We average across typically 1,000 Monte Carlo
representations to then allow a direct comparison between the
distribution of simulated and observed flux densities to be conducted.

\begin{figure}
    \centering
    \includegraphics[width=\columnwidth]{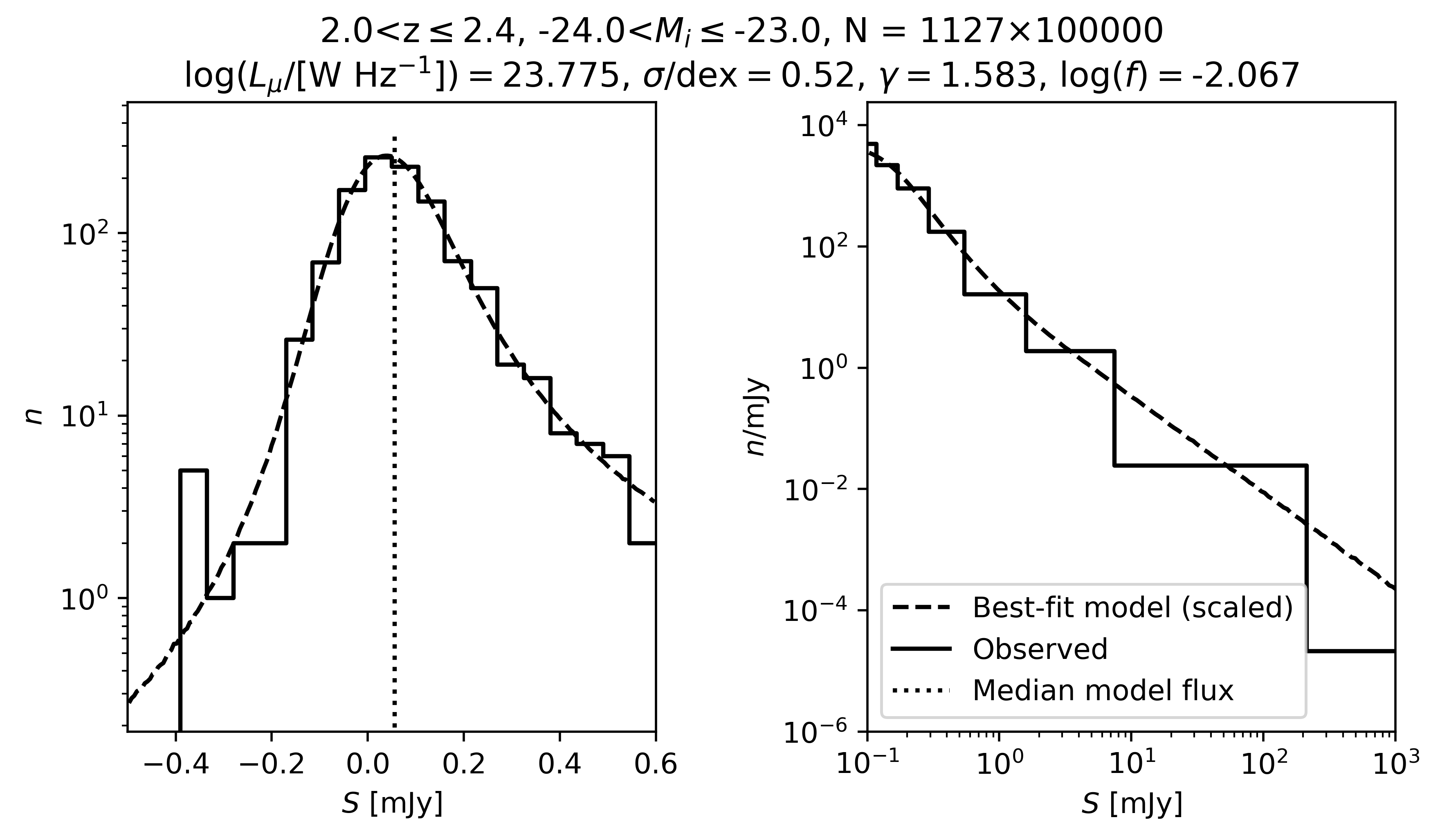}
    \caption{The observed flux density distribution for quasars with
      $2.0 < z \leq 2.4, -24 < M_i \leq -23$ (solid histogram) and the
      best-fit simulated flux density distribution for these (dashed
      line). The simulated distribution is generated by the model
      using the set of model parameters that provide the best fit to
      the observed flux density distribution for this subsample (see
      Section~\ref{Sec:ks}). This subsample hosts 1127 quasars and for
      each quasar, 100,000 representations of the model have been
      returned (and averaged). In order to clearly display both the
      low flux densities (with some noise-dominated negative values)
      and the tail to very high flux densities, the distribution is
      plotted in two parts. The left panel shows the observed flux
      density distribution at low flux densities, using a linear flux
      scale and equally-sized bins; the vertical dotted line indicates
      the median flux density, showing that this is offset from zero. The
      right panel shows the distribution at higher flux densities on a
      logarithmic flux scale, binned using the Bayesian Blocks
      \citep{BBlocks} formalism (note that the full unbinned
      distribution is used in the KS test). For the right panel, the
      count, $n$, has been divided by the width of the respective bin
      (therefore $n$/mJy) since the bins have varying width. Over all
      flux densities, the model provides an excellent match to the
      observed distribution. 
    \label{fig:flux_example}}
\end{figure}

An example of the comparison between the observed flux density
distribution for a given subsample (as defined in
Fig.~\ref{fig:L-z_plot}) and the best-fit simulated flux density
distribution for the same subsample can be seen in
Fig.~\ref{fig:flux_example}. The procedure to obtain the best-fit
model for a given observed flux densitiy distribution is detailed in
Section~\ref{Sec:ks}. It is clear that the model is able to provide a
good match to the observed flux density distribution. It is also
notable that the peak of both the observed and simulated distributions
are offset from 0\,mJy (this offset effectively constrains the SF
component of the model) and that even well below the 5-sigma noise
limit of the radio data (around 0.35\,mJy) the observed flux density
distribution is clearly non-Gaussian (with the tail of the
distribution to high flux densities constraining the jet contribution);
this highlights the valuable information available within the noise of
the radio data, and which is well-fitted by the model. It is worth
re-emphasising that our model assumes that no intrinsic bimodality
exists in the radio flux distribution of quasars, as illustrated in
Fig.~\ref{fig:flux_example}.

\subsection{Obtaining best-fit parameters} \label{Sec:ks}

The two-sample Kolmogorov-Smirnov (KS) test is a non-parametric
process to determine the probability that two sets of data have been
drawn from the same PDF. The test can therefore be used to compare the
simulated samples generated through the Monte Carlo realisations with
the observed radio flux density distributions. This
comparison will allow us to quantify the validity of our model as well
as to constrain the values of the free parameters.  Observed radio
flux density distributions are taken from the subsamples described in
Section~\ref{Sec:sample} and one example of these is illustrated in
Fig.~\ref{fig:flux_example}.

\begin{figure*}
    \centering
    \includegraphics[width=\textwidth,height=14cm]{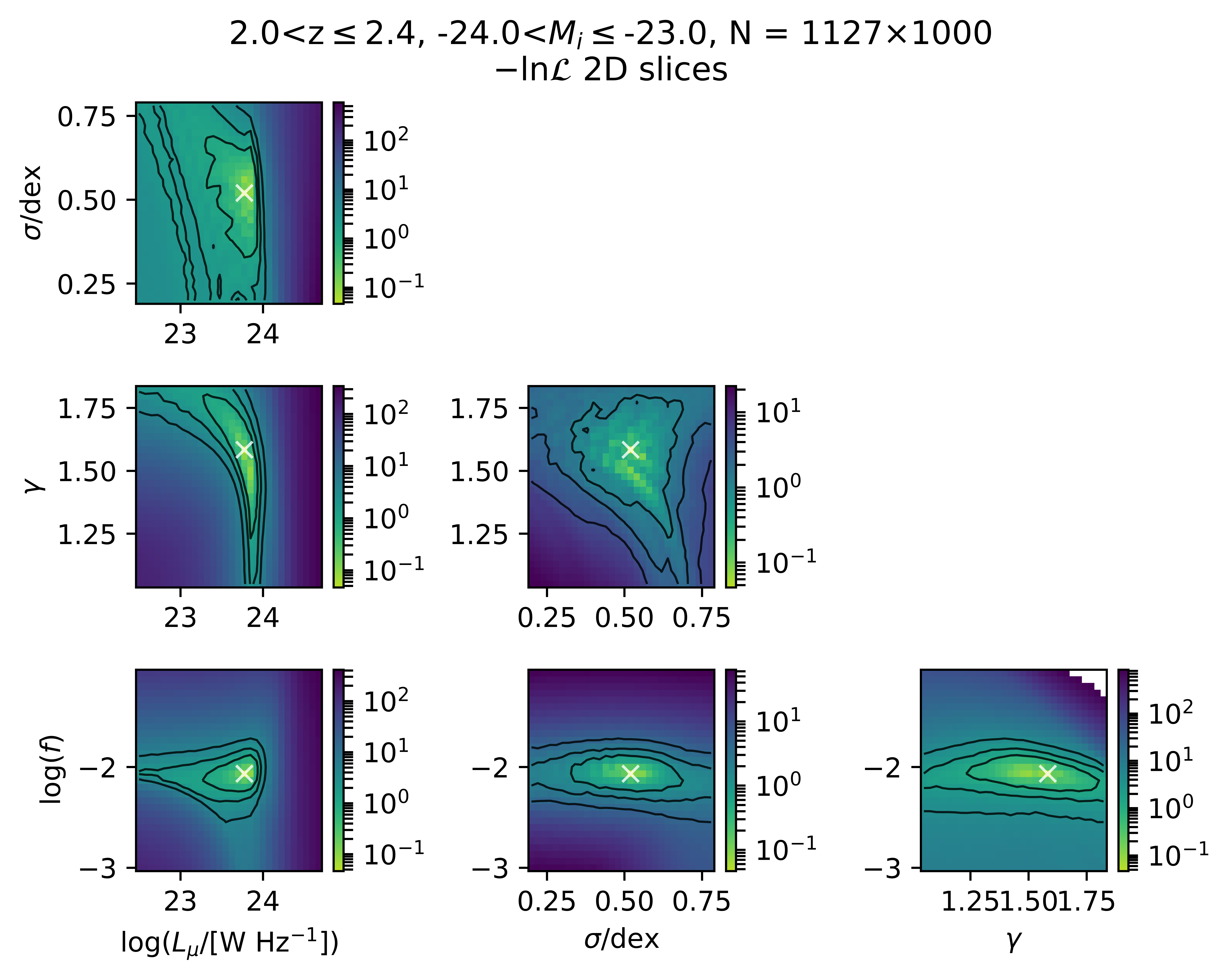}
    \caption{The log likelihood landscape for the given subsample
      ($2.0 < z \leq 2.4, -24 < M_i \leq -23$). 30 grid points in each
      dimension have been defined to compute $-\ln \mathcal{L}$. This
      subsample hosts 1127 quasars and for each quasar, 1000
      representations of the model have been returned (and
      averaged). The colourbar shows the value of the $-\ln
      \mathcal{L}$ computed, the white cross shows the best-fit values
      for the parameters and the black lines show the 1-,2-,3-sigma
      contours.}
    \label{fig:corner_example_30}
\end{figure*}

We found from initial tests that a simple KS test on the full flux
density distributions was not sensitive to the AGN component of the
model, and that the degeneracy between the normalisation, $f$, and the
slope, $\gamma$, was not being broken during our fitting. The reason
for this is the relatively few sources in the high luminosity tail
(the KS statistic is determined by the maximum absolute
  offset between the model and observed flux density distributions,
  and a tiny fractional difference in shape of the bulk of the
  population can lead to a bigger absolute offset in the cumulative
  distribution than that of a significant fractional offset in the
  tail of AGN).  This same issue prohibited a flux-binned chi-squared
approach.  Instead, to solve this problem, we split the flux density
distributions at a threshold of $S_{\rm thresh} = 1\mathrm{mJy}$. This
allows us to compare the relative number of sources above and below
the threshold using a chi-squared test (based on Poissonian
statistics; this ensures a broadly correct number of
  objects in the high luminosity tail), while additionally performing
the KS test separately on the flux density distributions either side
of the threshold; this approach gives additional weight
to the AGN component (which dominates the above-threshold
distribution). The value $S_{\rm thresh} = 1\mathrm{mJy}$ was chosen
as it is the flux density at which typically the observed flux density
distributions transition from the large peaked component to the tail
component of the population. In Appendix ~\ref{App:thresh} we
demonstrate that (to well within the errors) our results do not depend
on the choice of $S_{\rm thresh}$.

To optimise the parameters of the model, we extract the probabilities,
$p_>$ and $p_<$, returned from the KS tests above and below the flux
threshold, respectively. We then combine these with the chi-squared
value for the Poissonian number test, $\chi_{\rm P}^2$, to derive the
overall likelihood ($\mathcal{L}$)

\begin{equation} \label{eq:chisplit}
  -\ln \mathcal{L} = -\ln(P_>) -\ln(P_<) + \chi_{\rm P}^2/2
\end{equation}

\noindent which is then minimised. Here, 

\begin{equation}
    \chi_{\rm P}^2 = \frac{(N_{\rm model}^{<} - N_{\rm obs}^{<})^2}{N_{\rm model}^{<}}
    + \frac{(N_{\rm model}^{>} - N_{\rm obs}^{>})^2}{N_{\rm model}^{>}}.
\end{equation}

\noindent where $N_{\rm model}^{<}$ (or $N_{\rm model}^{>}$) is the
number of sources that the model predicts to lie below (or above) the
threshold and $N_{\rm obs}^{<}$ (or $N_{\rm obs}^{>}$) is the number
of sources observed below (or above) the threshold,

When finding the best-fit model for each of our subsamples, in order
to avoid the risk of minimisation routines getting stuck in local
minima (especially for less-populated, and hence noisier, regions of
parameter space), and given that the problem was computationally
manageable, we implemented a \textquote{brute force} method: we
computed the $-\ln \mathcal{L}$ value at every point on a multi-dimensional
parameter grid to find the global minimum and the uncertainties on
this. In each dimension of parameter space, 30 grid points were
defined, equally-spaced in the 4-dimensional parameter space:
$\mathrm{log}(L_{\mu}/[\mathrm{W\ Hz^{-1}}])$, $\sigma/\mathrm{dex}$,
$\gamma$, $\mathrm{log}(f)$. An example of the results of applying
such a minimisation routine can be seen in
Fig.~\ref{fig:corner_example_30}.

To ensure that the final best-fit model provides a valid explanation
of the data, we perform a single KS test between the overall modelled
flux density distribution and that of the observed quasars in each bin
in luminosity--redshift space. In all cases, the returned probability
is $\ge 0.25$, indicating that our two-component model for the radio
flux density distribution of quasars is valid across a wide range of
optical luminosities and redshifts.

\section{Results} \label{Sec:KSresults}

\subsection{Variation of model parameters with $M_i$ and $z$}
\label{Sec:params_withMi_z}

\begin{figure}
    \centering
    \includegraphics[width=\columnwidth]{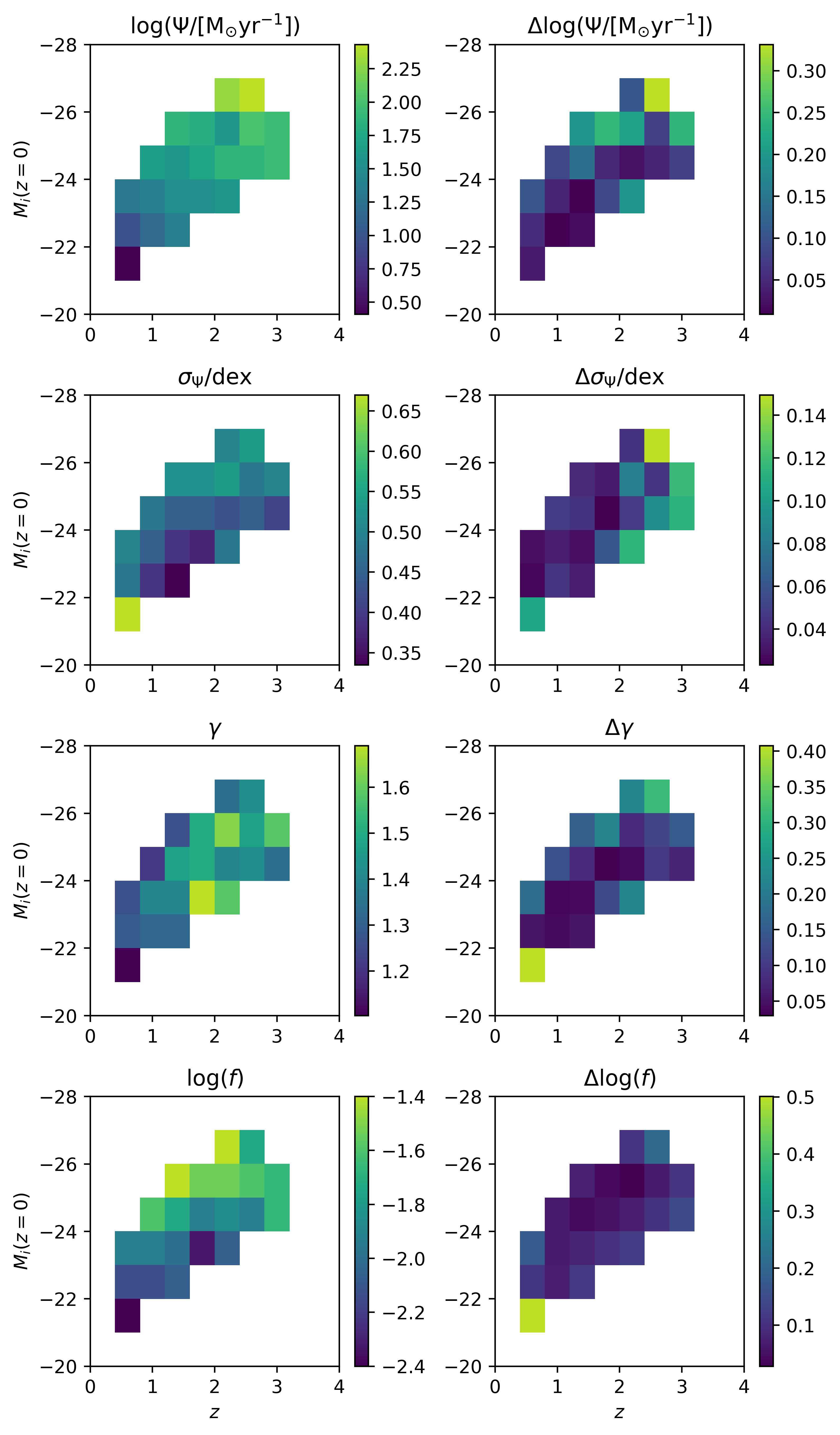}
    \caption{2D distribution (in $M_i - z$ space) of the best-fit
      values for the four model parameters (left) and their 1$\sigma$
      errors (right) from the Monte Carlo simulations, obtained
      through minimisation of $-\ln \mathcal{L}$. }
    \label{fig:KS_grid}
\end{figure}

Given that the model seems to accurately reproduce the observed data,
we can now look at how the best-fit parameters change with optical
luminosity and redshift. The best-fit parameters along with their
errors for the subsamples are visualised in $M_i - z$ space in
Fig.~\ref{fig:KS_grid}. For clarity, the collapsed version of these
results is given in Fig.~\ref{fig:KS_grid_collapse}, and a full table
of best-fit parameters is provided in Appendix~\ref{App:tabs}
(Table~\ref{Tab:allsample}). The parameters $L_{\mu}$ and $\sigma$
that characterise the Gaussian component of the model are expressed
respectively as their SF equivalents, $\Psi$ and $\sigma_{\Psi}$,
through application of the calibration given in
Section~\ref{Sec:SFcomp}.

One immediate result of note is the lack of discernible trends, with
either redshift or optical luminosity, in the width of the SF
component and the slope of the AGN component. These findings are
perhaps expected. That the scatter in the relationship between SFR and
quasar luminosity does not change much with quasar luminosity has been
seen before in literature plots \citep[e.g.][]{Lanzuisi2017}, and as
this ratio is driven by gas distributions within the host galaxies it
is not surprising that there is no strong redshift dependence either.
The slopes of the radio luminosity function are also known to not
evolve significantly with redshift \citep[e.g.][and
    references therein]{DP90,Smolcic2017}. Given the lack of
observed trends of these parameters, we took the decision to fix
$\sigma$ and $\gamma$ and conduct further analysis of the remaining
two parameters. Fixing $\sigma$ and $\gamma$ allows us to increase the
resolution with which we fit the remaining parameters as well as
removing the possibility of artificial jumps in the best-fit
parameters for $L_{\mu}$ and $f$ caused by random variations in
$\sigma$ and $\gamma$. We therefore reserve discussion of the trends
of the mean SFR rate and the jet power normalisation to
Section~\ref{Sec:fix}.

\begin{figure}
    \centering
    \includegraphics[width=\columnwidth]{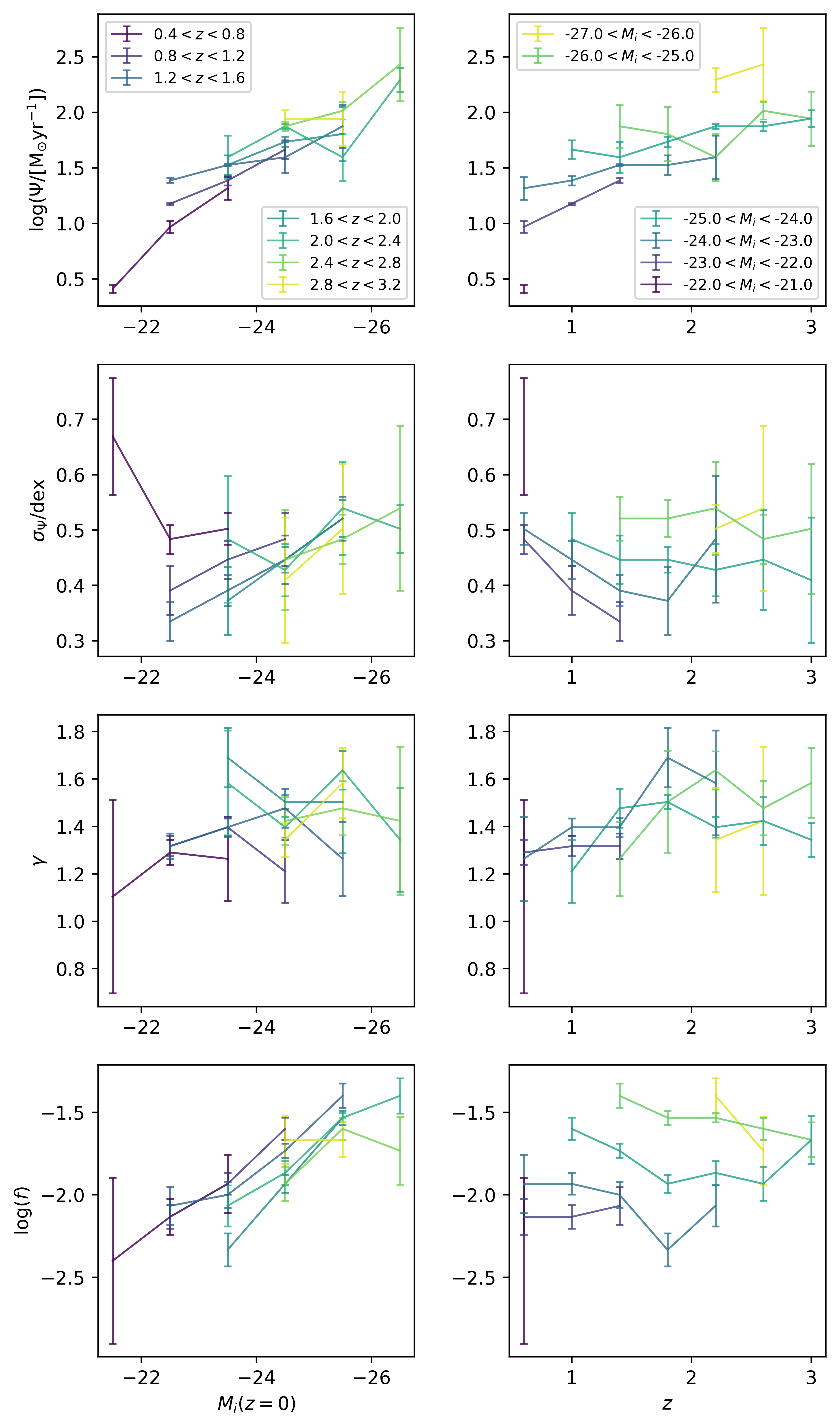}
    \caption{1D collapsed distribution showing the variation of the
      best-fit values for the four model parameters (top to bottom) as
      a function of the optical luminosity of the quasar (left) and
      redshift (right). The results indicate that the width of the SF
      Gaussian distribution ($\sigma$) and the slope of the jet power
      distribution ($\gamma$) are broadly constant with both optical
      luminosity and redshift. The SFR ($\Psi$) increases strongly
      with optical quasar luminosity, and more weakly with redshift,
      while the jet-power normalisation ($f$) increases with optical
      luminosity but is independent of redshift.}
    \label{fig:KS_grid_collapse}
\end{figure}

\subsection{Fixing $\sigma$ \& $\gamma$}
\label{Sec:fix}

We fix the width of the Gaussian and the slope of the power-law with
the values $\sigma = 0.45$ (corresponding to $\sigma_{\Psi}=0.42$ from
Equation~\ref{eqn:sfr_l150}) and $\gamma = 1.4$; these
  are the (suitably-rounded) weighted-mean values from the different
  grid cell fits. We also increase the resolution of our parameter
grid such that the number of grid points in each of the two remaining
dimensions is now 40.

\begin{figure}
    \centering
    \includegraphics[width=\columnwidth]{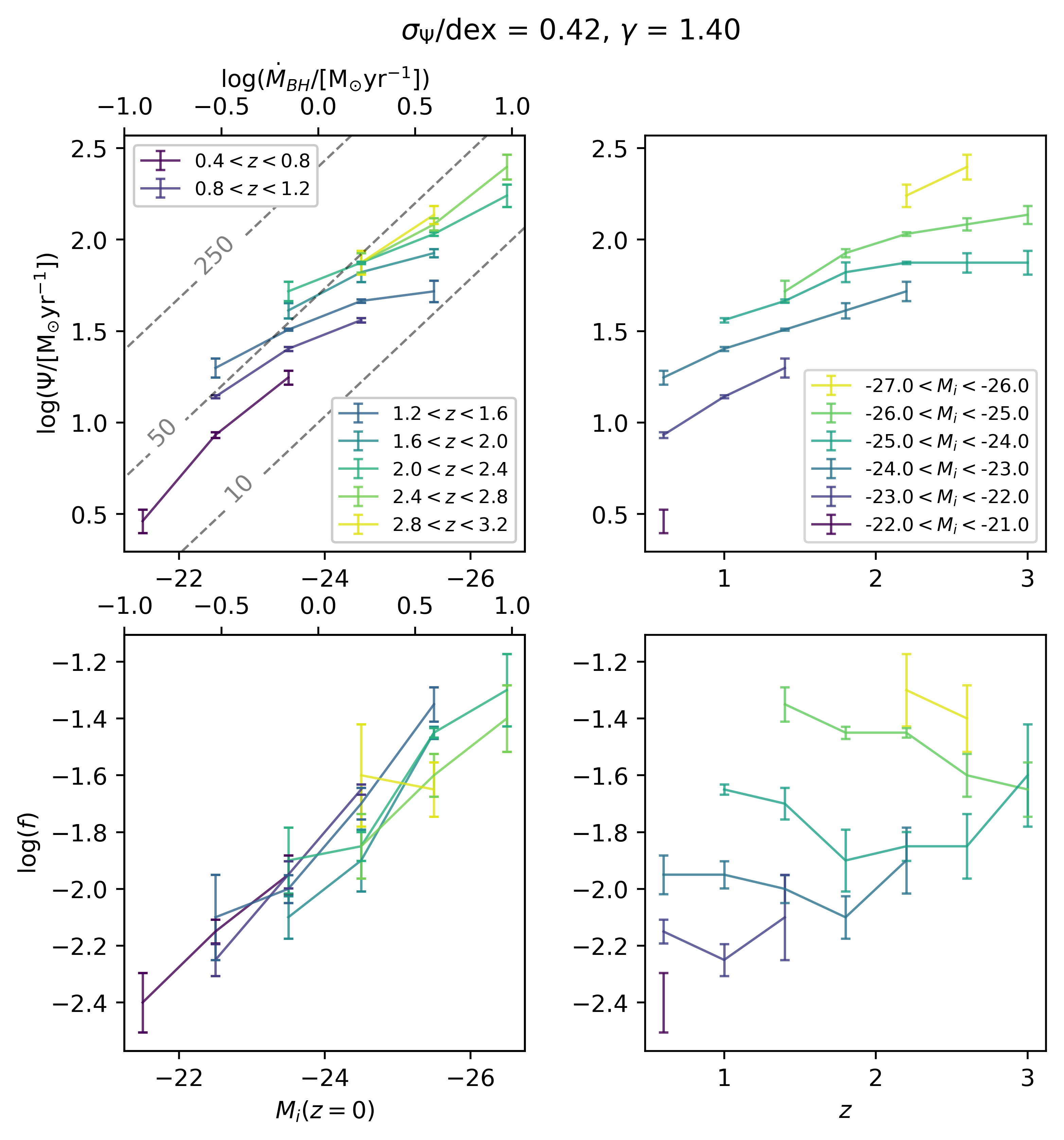}
    \caption{The variation of the SFR ($\Psi$; top) and the jet power
      normalisation ($f$; bottom) with optical quasar luminosity
      (left) and redshift (right) from the Monte Carlo simulations
      with fixed $\sigma,\ \gamma$. The upper-left panel shows that the
      SFR increases with increasing optical luminosity (i.e.\ BH growth
      rate); the dashed lines indicate ratios of $\Psi/\dot{M}_{\rm
        BH}$ of 10, 50 and 250. The upper-right panel shows that the
      SFR of the quasar hosts increases with redshift out to redshift
      $z \sim 2$ and then flattens. The lower panels show that the jet
      power normalisation correlates strongly with optical luminosity
      but shows no consistent trend with redshift.}
    \label{fig:KS_grid_marg_fix}
\end{figure}

\begin{figure*}
  \includegraphics[width=0.9\textwidth,height=12cm]{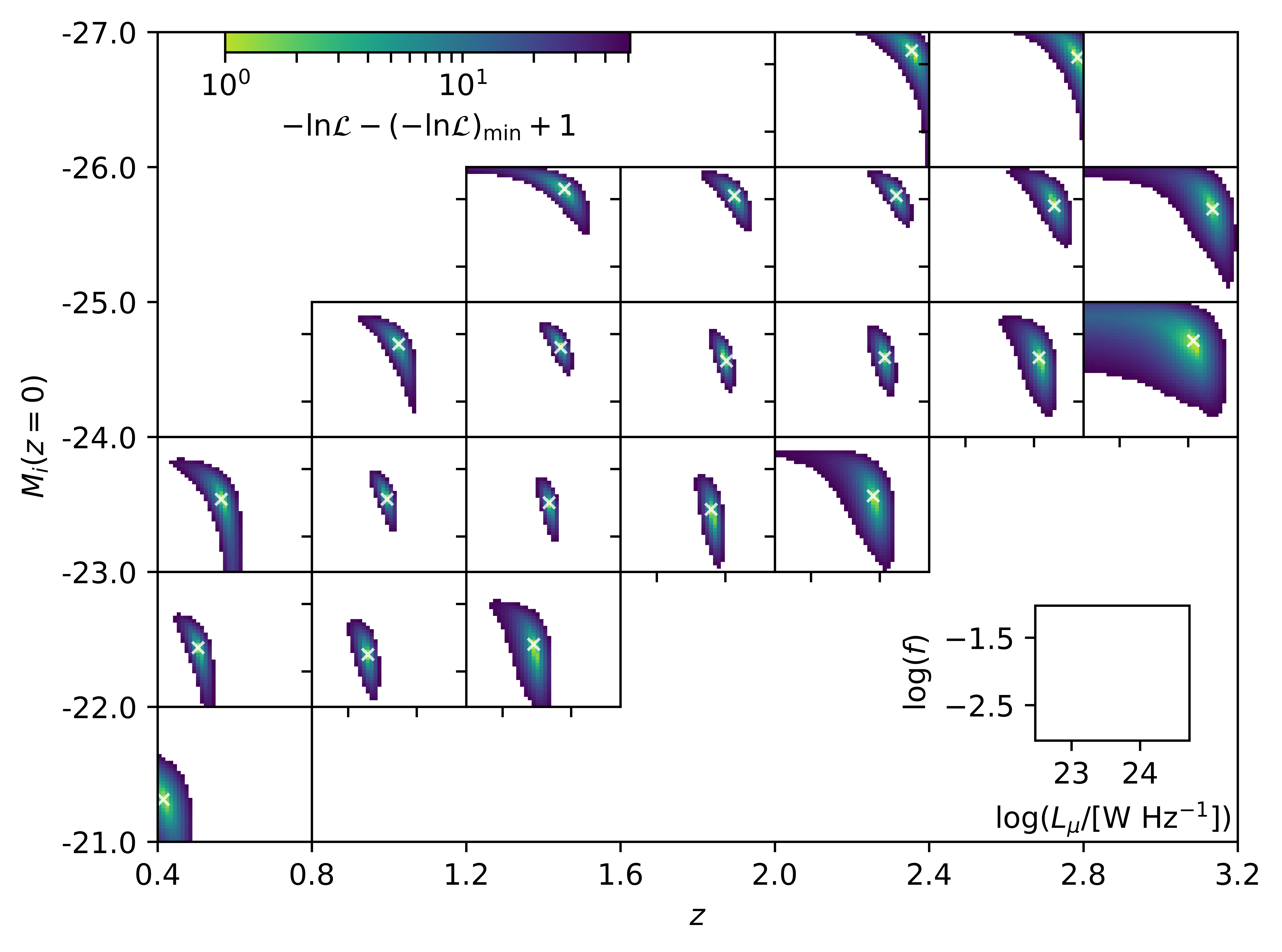}
  \caption{The distribution of log likelihood across the parameter
    space ($L_{\mu}, f$) explored for each subsample in $M_i - z$
    space. The colourbar shows the change in $-\ln \mathcal{L}$ from the
    minimum of each respective block (where the minimum is set to 1
    since we are using a logarithmic colourbar). The white cross shows
    the best-fit values of the model parameters for that given
    subsample. The lower-right inset indicates the ($L_{\mu}, f$) axis
    ranges for each of the distributions. The set of plots allow easy
    visualisation of how the best-fit values of $L_{\mu}$ and $f$ vary
    across luminosity-redshift space.
    \label{fig:distplots}}
\end{figure*}

Best-fit parameter values are provided in Table~\ref{Tab:allsample}.
The collapsed distributions of the best-fit parameters with $M_i$ and
$z$ are shown in Fig.~\ref{fig:KS_grid_marg_fix}, while the full
$-\ln\mathcal{L}$ parameter space is visualised in Fig.~\ref{fig:distplots}.
In Fig.~\ref{fig:KS_grid_marg_fix} we also relate $M_i$ to an estimate
of the growth rate of the SMBH, $\dot{M}_{\rm BH}$, as described in
Section~\ref{Sec:accretion}, as it provides some interesting physical
information; this is shown as an upper x-axis. Reassuringly, the
results with fixed $\sigma$ and $\gamma$ are in agreement with the
trends observed in Fig.~\ref{fig:KS_grid_collapse} when these two
parameters were not fixed, but now with lower noise.

The SFR, $\Psi$, of the host is seen to increase with increasing
optical luminosity (or SMBH growth rate). We also find that the SFR of
the host increases with increasing redshift out to $z\sim2-3$ at which
point, we observe the SFR beginning to turnover. In
Appendix~\ref{App:binning} we confirm that this redshift trend is
genuine and is not caused by selection effects related to correlations
between $M_i$ and $z$ within each bin.

We also find a strong dependence of the jet power normalisation with
optical luminosity, in line with previous studies
\citep[e.g.][]{Jiang2007}: at the lowest optical luminosities, we
observe values for the fraction of sources at high radio luminosities
($L_{\rm 150 MHz} > 10^{26}$W\,Hz$^{-1}$), $f\leq0.5\%$ while at the
highest optical luminosities, we reach values of $f\sim6\%$ (see
Figures~\ref{fig:KS_grid_marg_fix} and \ref{fig:distplots}). No
discernible trend of $f$ with redshift exists, suggesting that the
principal mechanism that drives the distribution of jet powers must be
an internal property of the system.

\subsection{Black Hole Mass Dependence}
\label{Sec:BHmass_analysis}

\begin{figure*}
    \includegraphics[width=13cm]{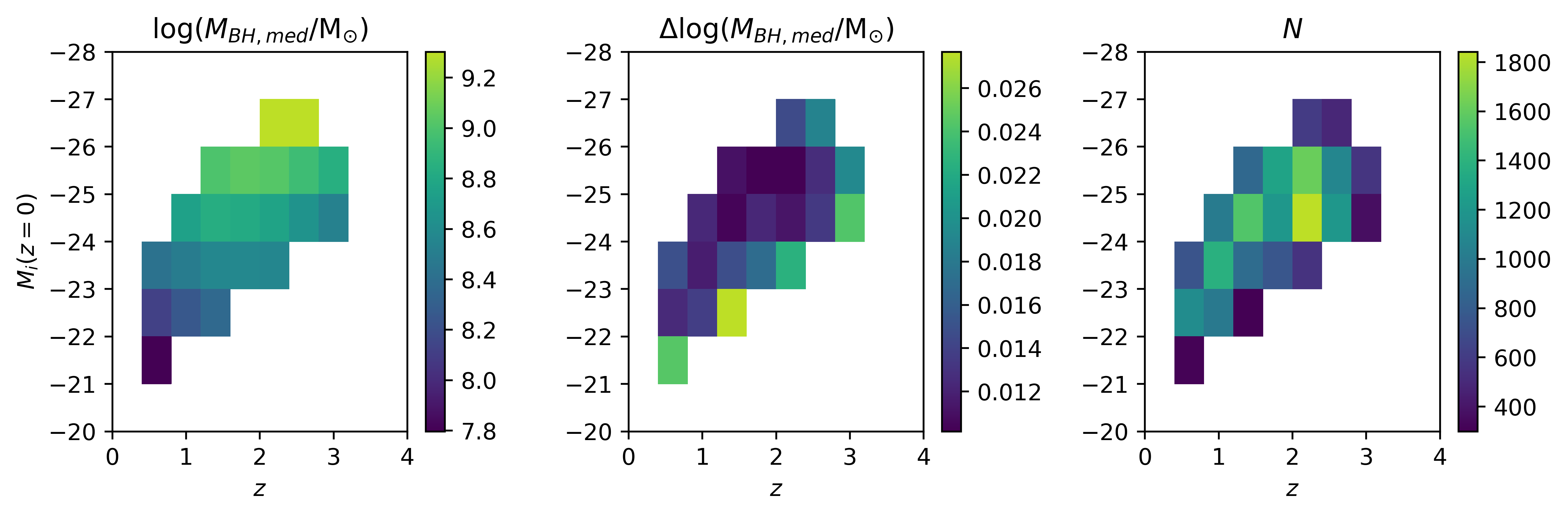}
    \caption{From left to right: (i) The median virial BH mass
      estimate in each subsample of the $M_i - z$ grid, used to divide
      each subsample into two. (ii) The error on the median BH mass,
      calculated using the typical error on the median ($\Delta
      \mathrm{log}(M_{\rm BH,med}) = 1.253
      \frac{\sigma_{\mathrm{log}M_{BH}}}{\sqrt{N}}$ where
      $\sigma_{\mathrm{log}M_{BH}}$ is the standard deviation of the
      distribution of log BH mass, and $N$ is the number of quasars in the
      subsample). (iii) The number of quasars, $N$, in the subsample
      that have a BH mass estimate.}
    \label{fig:mbhmed_grid}
\end{figure*}

We investigate the effect of the BH mass on the model parameters, for
the case of fixed $\sigma = 0.45,\ \gamma=1.4$, using only those
sources that have a BH mass estimate, as described in
Section~\ref{Sec:BHmass}. To investigate the effect of BH mass on the
model parameters, we split each subsample in $M_i - z$ space in two at
the median BH mass of the given subsample, $M_{\rm BH,med}$. The
median BH mass of each subsample was chosen rather than a uniform
value for the entire sample as the BH mass varies across $M_i - z$
space, as can be seen from the first panel of
Fig.~\ref{fig:mbhmed_grid}. Across most of the parameter space, the
difference in the median BH mass between the higher and lower BH mass
bin is a factor of 2.5-3.5.

\begin{figure}
    \centering
    \includegraphics[width=\columnwidth]{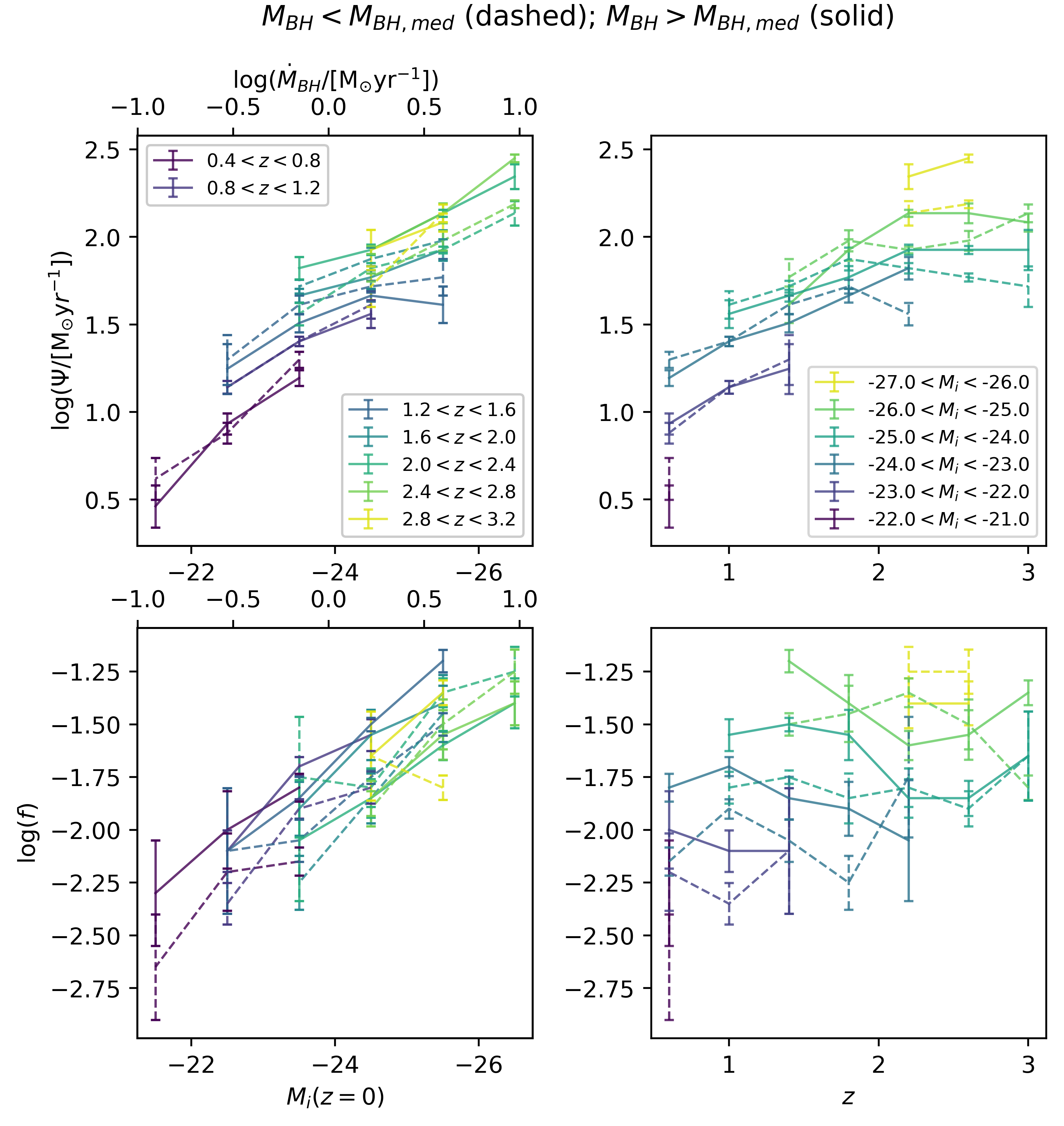}
    \caption{The variation of the SFR ($\Psi$; top) and the jet power
      normalisation ($f$; bottom) with optical quasar luminosity
      (left) and redshift (right), split into bins of higher (solid
      lines) and lower (dashed lines) black hole mass, as outlined in
      the text. We observe little difference in the SFR between the
      two BH mass bins, except perhaps at $z>2$ where we see hints of higher
      SFRs in the higher BH mass bin. For the jet power normalisation,
      $f$, we find this is typically a factor $\sim 2$ higher in the
      higher BH mass bin, at least for $z<2$.}
      \label{fig:KS_grid_marg_fix_bh}
\end{figure}

The results of including the BH mass information can be seen in
Fig.~\ref{fig:KS_grid_marg_fix_bh}; see also Table~\ref{Tab:BHmass}.
The general trends of the parameters of the model are still in
agreement with those found and discussed in Section~\ref{Sec:fix} and
so will not be repeated. Instead, we focus on how the inclusion of the
virial BH mass estimate affects the trends. We find little variation
of the SFR trend with optical luminosity between the high and low BH
mass bins. Similarly, at $z<2$, the SFR seems to be independent of the
BH mass. However, at redshifts above $z\approx2$ (where the SFR
flattens) in almost all luminosity-redshift bins the SFR in the higher
BH mass bin is higher than that at lower black hole masses, by an
average of 0.17 dex. Although the significance of this is low
(typically 1-2$\sigma$) in each case, the consistent direction of the
offset across almost all of the high-redshift bins suggests a genuine
effect, although more data would be required to confirm this.

Regarding the fraction of sources at high radio luminosities, for
quasars with $z<2$ there are indications that an increase in BH mass
boosts the value of $f$: although once again the differences are of
relatively low significance (typically 1-2$\sigma$) within each
individual luminosity--redshift bin, for all 13 luminosity--redshift
bins at $z<2$ the value of $f$ derived for the higher black hole mass
subsample is greater than or equal to that of the lower black hole
mass subsample (giving a much higher combined
  significance). The average difference is a factor of $\sim 2$ in
$f$ (see Table~\ref{Tab:BHmass}). This result suggests that, at $z<2$,
the jet power normalisation may depend on both $M_{\rm
  BH}$ and $M_i$ (or $\dot{M}_{\rm BH}$), independently. At higher
redshift, $z>2$, there is no evidence for a significant black hole
mass dependence of $f$.

\section{Discussion \& Physical Interpretations}
\label{Sec:discussion}

Before we discuss the physical interpretations of our results, it is
worth quickly summarising what we have found.

\begin{itemize}
    \item The radio emission of quasars can be explained by our simple
      model, which assumes two sources of radio emission contribute in
      all quasars: SF in the host galaxy and the AGN. The SF in the
      host galaxy is modelled as being drawn randomly from a Gaussian
      distribution of given centre and width. The AGN (jet) luminosity
      is drawn randomly from a power-law distribution of given
      normalisation and slope. The four parameters of the model have
      been constrained and their dependencies on redshift, optical
      luminosity and BH mass have been investigated.
    \item We find that the width of the Gaussian SF component and the
      slope of the power-law AGN component do not vary significantly
      with either redshift or optical luminosity. This
      is perhaps expected, as explained in
      Section~\ref{Sec:params_withMi_z}, justifying our approach to
      fix the two parameters during further analysis. This allowed us
      to increase the resolution and clarity with which we constrain the
      remaining two parameters.
    \item We observe that the SFR of quasar host galaxies increases
      with redshift out to $z\sim2$ and then flattens in the range
      $z\sim2-3$. Independently, the SFR also increases strongly with
      the quasar luminosity. We observe little dependence of the SFR
      on BH mass until we get to the highest redshifts, at which point
      we see some evidence that increasing the BH mass may correspond
      to a small increase in the SFR.
    \item The normalisation of the jet power (and hence the fraction
      of sources at high radio luminosities) shows very little trend
      with redshift. However, it increases with
        increasing optical luminosity and there are indications
        that it may also increase independently with the SMBH mass.
\end{itemize}

\subsection{The ubiquity of radio jets}
\label{Sec:disc_bimodal}

We find that our two-component prescription of the radio emission, SF
and AGN, is valid in describing the observed radio flux density
distribution of quasars across a wide range of redshifts and optical
luminosities.  As our model inherently assumes a wide and continuous
distribution of radio jet powers, we therefore argue that the
historical dichotomy of RQ and RL quasars can be simply explained by
whether the radio emission of the quasar is dominated primarily by the
SF of the host galaxy or the powerful large-scale jets.

Given the validity of our model, this implies that both jets and SF
are contributing in all quasars, as argued by \citet{Gurkan2019}.
Since quasars typically reside in star-forming galaxies, we do expect
a contribution at some level from SF. However, the ubiquity of quasar
jets is not necessarily expected. It should be emphasized that such
ubiquity is not conclusively demonstrated by our analysis: when SF
dominates the radio emission, our flux density distribution comparison
cannot distinguish between the scenario where some quasars have a very
weak jet contribution that lies significantly below the SF
contribution (the lower limit of the power-law distribution from which
jet luminosities are drawn is typically in the range
$\mathrm{log}(L_l/[\mathrm{W\ Hz^{-1}}]) \approx 20 -22$, whereas the
mean of the SF Gaussian distribution is typically
$\mathrm{log}(L_{\mu}/[\mathrm{W\ Hz^{-1}}]) \approx 22.5 - 24.5$) and
the scenario where some quasars have no jet contribution to the radio
emission (e.g. if there is a threshold for jet production). However,
our results do demonstrate that radio emission from quasar jets is
contributing at least down to luminosities where the radio emission
from SF becomes comparable, which is well into the traditionally
`radio-quiet' quasar regime. This is consistent with the high
detection rates of emission from small-scale low-luminosity jet-like
structures in deep, high angular resolution radio observations of
some radio-quiet quasars
\citep[e.g.][]{Ruiz2016,Jarvis2019,Hartley2019}, and with the presence
of radio jets with luminosities down to at least $10^{21}$W\,Hz$^{-1}$
in massive galaxies that do not host quasars
\citep[e.g.][]{Sabater2019,Mingo2019,Baldi2021}.

\begin{figure}
    \centering
    \includegraphics[width=\columnwidth]{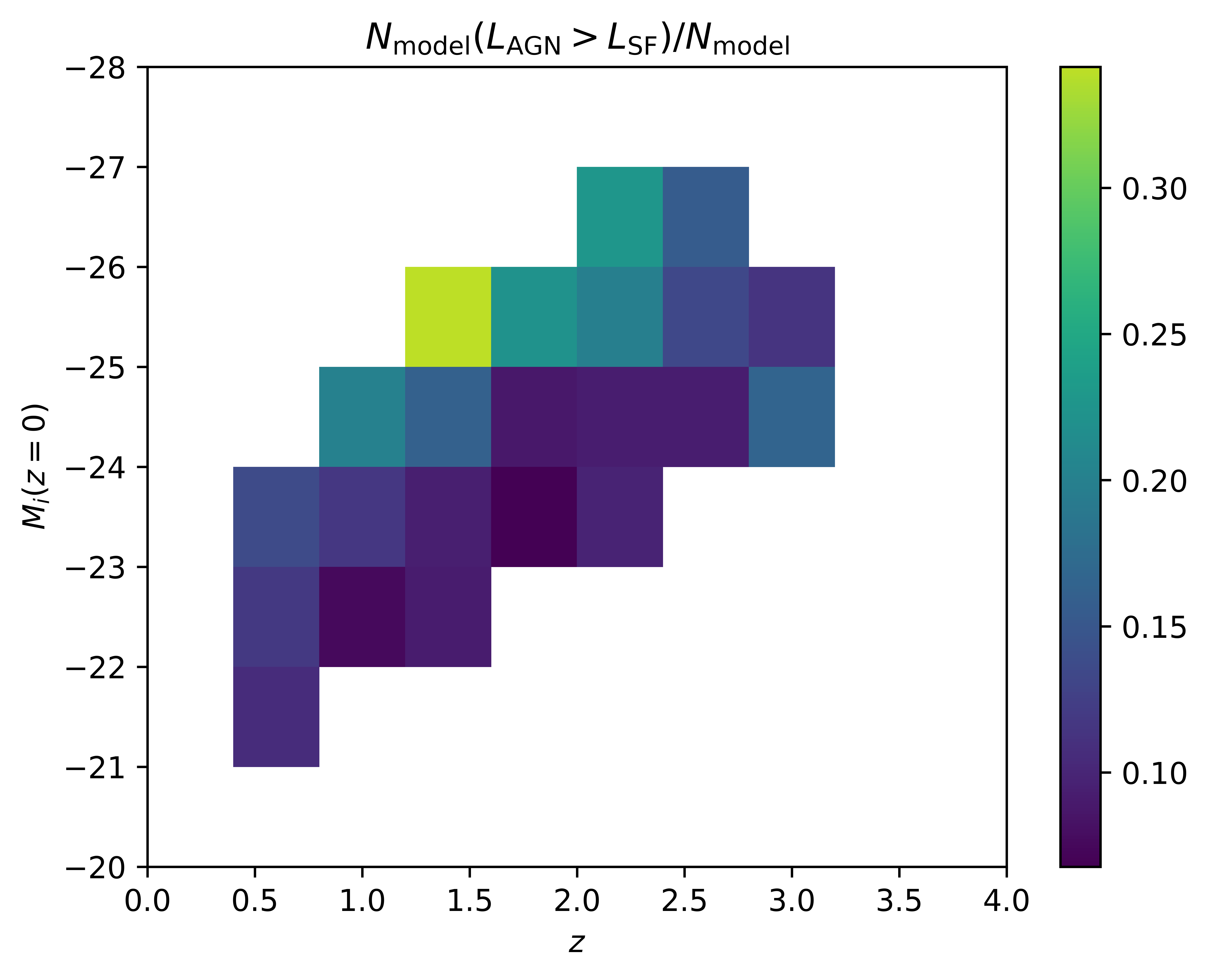}
    \caption{The fraction of quasars in our sample for which the
      contribution to the radio emission from a jet component exceeds
      that from star formation, as a function of redshift and quasar
      luminosity. The 10-20\% values derived match typically-quoted
      values for the radio-loud fraction. We find that the radio-loud
      fraction is typically higher for higher quasar luminosities but
      shows no uniform trend with redshift.
      \label{fig:RL_fraction}}
\end{figure}

In Fig.~\ref{fig:RL_fraction} we present the fraction of
quasars for which the jet contribution to the radio emission is larger
than that of the star-formation contribution in our model, as a
function of redshift and quasar luminosity. The derived values of
10--20\% agree well with the widely-adopted `radio-loud fraction' for
quasars \citep[e.g.][]{Kellermann1989}.  We find a clear trend for an
increasing jet-dominance with increasing optical luminosity, from of
order 10\% at optical luminosities $M_i > -24$ to $\sim 20$\% at $M_i
< -25$. This is in line with most literature results for the
dependence of radio-loud fraction on quasar luminosity
\citep[e.g.][]{Jiang2007}. Overall we find no uniform trend for how
the jet-dominated fraction varies with redshift. At high luminosities
($M_i < -25$) there is indication for a decreasing jet-dominated
fraction with increasing redshift, as suggested by \citet{Jiang2007},
but at lower luminosities any redshift trend is weaker, absent or
non-monotonic. This lack of redshift evolution is in line with recent
studies out to the highest redshifts \citep[e.g.][and references
  therein]{Liu2020}.

\begin{figure}
    \centering
    \includegraphics[width=\columnwidth]{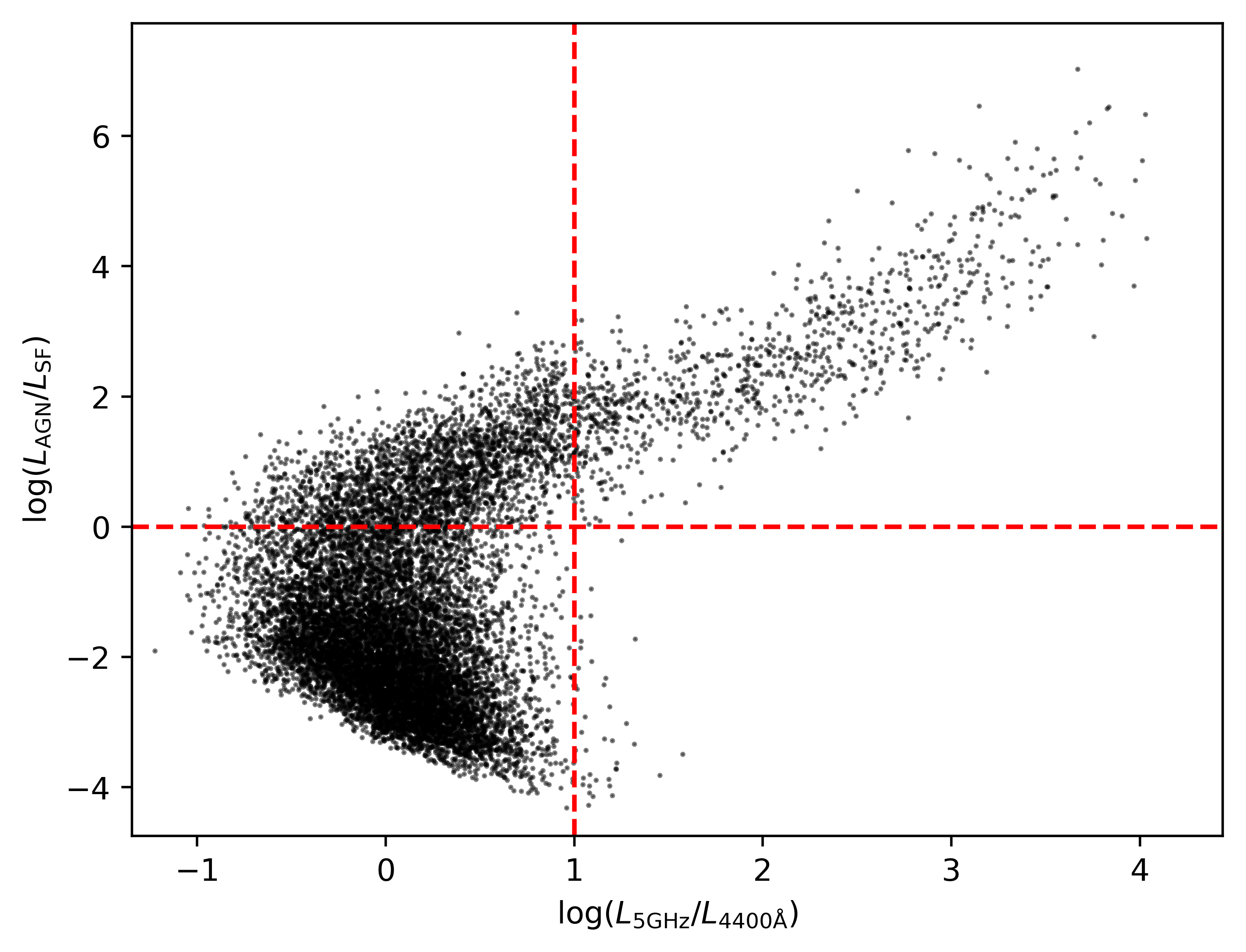}
    \caption{A comparison between the traditional
        `radio-loud' definition based on $R = L_{\rm 5GHz} / L_{\rm
          4400\text{\normalfont\AA}}$ (with a cut-off value at $R=10$)
        and the modelled $L_{\rm AGN} / L_{\rm SF}$ ratios, for all
        quasars detected with $S/N>2$ in LoTSS (sources with limits
        would fill out the lower-left region; see
        Sec.~\ref{Sec:disc_bimodal} for more details). The traditional
        cut-off is seen to select a clean sample of quasars which are
        highly jet-dominated; however, a number of sources that do not
        satisfy the traditional radio-loud cut also have their radio
        emission dominated by the radio jets.
      \label{fig:RL_comparison}}
\end{figure}

It is interesting to compare these models in detail with traditional
techniques of selecting radio-loud AGN. The traditional radio-loudness
parameter, $R=L_{\rm 5GHz} / L_{\rm 4400\text{\normalfont\AA}}$, can
be calculated for these quasars using the LoTSS data and an assumed
spectral index of $\alpha = 0.7$; in all of the redshift {\it vs}
absolute magnitude space studied, the LoTSS data are sensitive enough
to probe down to below the usual $R=10$ cut-off value. The models
do not predict the jet and star-formation components for
  each individual quasar, but rather predict the distribution of
properties for a population of quasars in each luminosity-redshift
grid cell. To allow a direct comparison, we therefore consider a
single iteration of the model in each grid cell, rank the modelled
sources by total radio luminosity, and cross-match these against the
observed data in the same grid cell, similarly ranked by radio
luminosity. Figure~\ref{fig:RL_comparison} compares the observed value
of $R=L_{\rm 5GHz} / L_{{\rm 4400}\text{\normalfont\AA}}$ for each
quasar against the modelled $L_{\rm AGN} / L_{\rm SF}$ for its
rank-matched model source. It is important to note that
  the ranking process adopted is only viable for those sources
  detected by LoTSS (with signal-to-noise above 2), and that therefore
  sources with upper limits on their radio luminosities are not shown
  on the plot. These sources would fill out the lower left region of
  the plot; the detection limits (and typical $L_{\rm AGN} / L_{\rm
    SF}$ ratios) are different in each grid cell in redshift
  and optical luminosity, and this produces the apparent shape of the
  cut-off towards the lower-left (which is artificial).

As can be seen from Figure~\ref{fig:RL_comparison}, the traditional
radio-loud definition of $R = L_{\rm 5GHz} / L_{{\rm
    4400}\text{\normalfont\AA}} > 10$ cleanly selects a population of
quasars for which the jet is by far the dominant source of radio
emission (typically $L_{\rm AGN} / L_{\rm SF} > 10$); the cut-off
value of $R=10$ is shown to be well-motivated as cuts at lower values
of $R$ would begin to pick up a population of quasars whose radio
emission is dominated by star-formation (quasars with $R > 1$ and
$L_{\rm AGN} \ll L_{\rm SF}$ are objects where the star-formation is
strong enough to give significant radio emission; typically these are
lower-redshift, lower optical lumoinosity quasars). On the other hand,
it is also clear from Figure~\ref{fig:RL_comparison} that there is a
significant population of sources with $L_{\rm 5GHz} / L_{{\rm
    4400}\text{\normalfont\AA}} < 10$, that would hence
be traditionally classified as radio-quiet, for which the jet is still
the dominant source of the radio emission. For some of
  these `radio-quiet' sources, the jet produces two orders of
  magnitude more radio emission than the star-formation.

\begin{figure}
    \centering
    \includegraphics[width=\columnwidth]{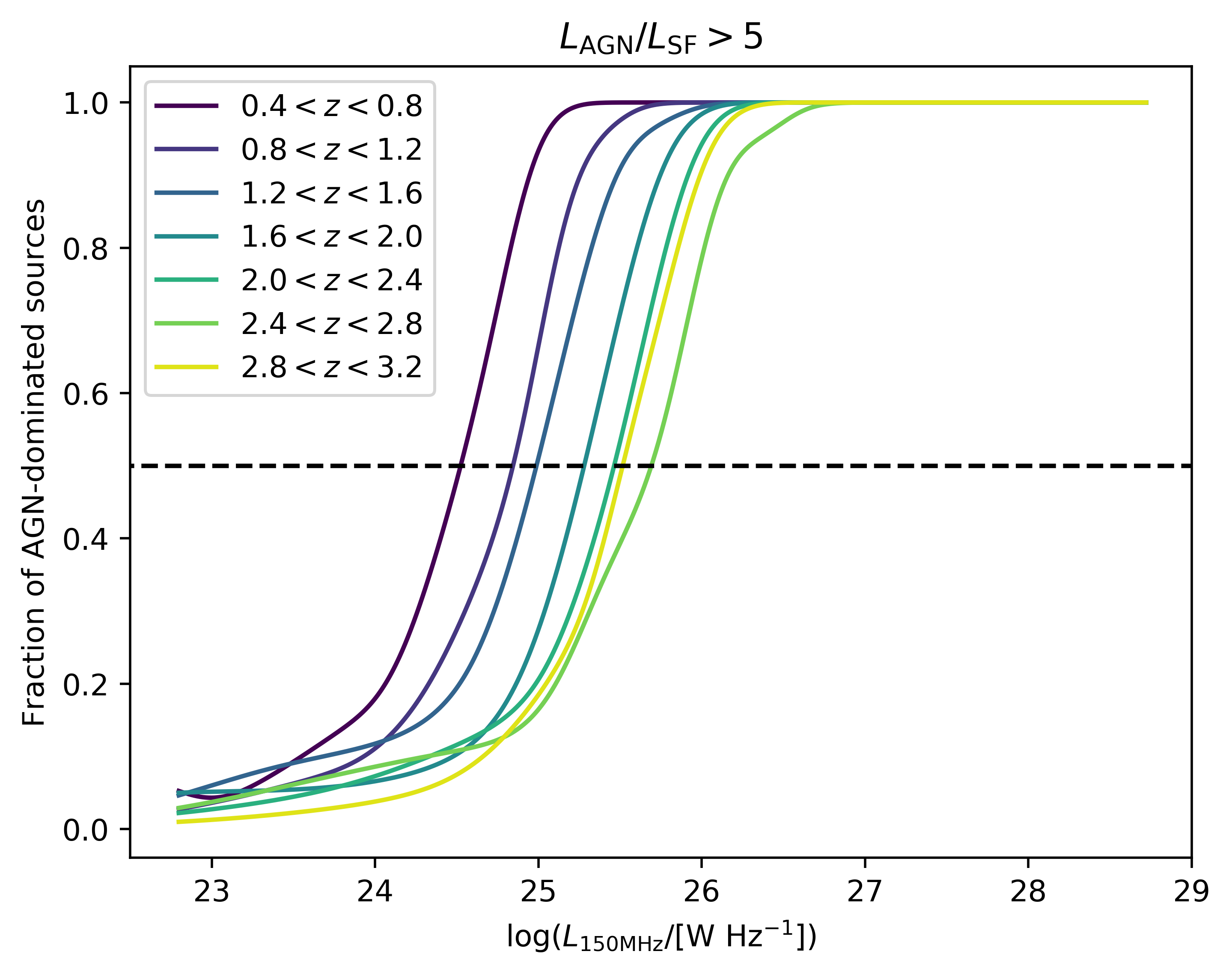}
    \caption{The fraction of radio sources at each
        redshift for which the jet is the dominant source of the radio
        emission ($L_{\rm AGN} / L_{\rm SF} > 5)$, as a function of
        radio luminosity. The plot is created by convolving the radio
        luminosities with a kernel density estimator of width 0.2
        dex. The transition of the population from
        star-formation-dominated to jet-dominated happens over an
        order of magnitude in radio luminosity in each case, and the
        mid-point radio luminosity depends strongly on redshift
        (either directly, or indirectly due to the correlation between
        redshift and absolute magnitude). A radio-loudness
        classification based solely on radio luminosity would not
        produce a representative sample of jet-dominated sources.}
      \label{fig:RL_comp2}
\end{figure}

The other definition of radio-loudness widely used in the
  literature is a selection purely on the basis of radio
  luminosity. Figure~\ref{fig:RL_comp2} shows the fraction of sources
  for which the jet is significantly the dominant source of radio
  emission ($L_{\rm AGN} / L_{\rm SF} > 5$), as a function of radio
  luminosity, in different bins of redshift. In each redshift bin, the
  transition from less than 10\% of sources satisfying this criteria,
  to over 90\% satisfying it, happens over typically an order of
  magnitude range in radio luminosity. Furthermore, the radio
  luminosity at which 50\% of sources satisfy the criteria increases
  with redshift; this is likely to be at least partially driven by the
  increase in quasar optical luminosity with redshift due to the
  sample selection effects. It is clear that a simple radio luminosity
  cut offers only a very crude manner of selecting jet-dominated
  sources.

\subsection{The star-formation rates of quasar host galaxies}
\label{Sec:disc_SFRs}

From analysis of the SF component of our model, we find that the
typical SFR increases with increasing optical luminosity of the
quasar. Intuitively, this appears reasonable: if there is more gas
available in the galaxy then there will be both more for the BH to
accrete (higher optical quasar luminosity) and more available to form
stars, in line with the Kennicutt-Schmidt Law, $\sum_{\rm
  SFR}\propto\sum_{\rm gas}^{1.4\pm0.15}$ \citep{Kennicutt1998}. We
also find that the SFR of the quasars increases with redshift out to
at least $z\sim2$. These trends of the SFR of the quasar host with
redshift and optical luminosity are seen independently of each other.

The evolution of the SFR of the quasar hosts with redshift is in line
with many results in the literature which find the same increase
\citep[e.g.][]{Bonfield2011,Rosario2012,Mullaney2012,Harrison2012}. This
is likely to be related to more gas being available at earlier times.
However, the strong correlation of SFR with optical luminosity of the
quasar appears, at first glance, to be at variance with recent studies
that have concluded that, once redshift effects are accounted for, any
correlation of SFR with quasar luminosity is either weak
\citep[e.g.][]{Harrison2012,Azadi2015,Stanley2017,Stemo2019} or
insignificant \citep[e.g.][]{Rosario2012,Mullaney2012,Stanley2015}.
However, these studies have been typically based on moderate
luminosity AGN ($L_{\rm bol}$ in the range $10^{35} - 10^{38}$W),
selected through deep X-ray observations in relatively small
fields. The AGN in the current study are rarer, higher-luminosity
quasars ($L_{\rm bol} \sim 10^{38}-10^{40}$W): previous studies which
have probed to these high AGN luminosities find, like our study, that
SFR does correlate with AGN luminosity in this regime
\citep[e.g.][]{Shao2010,Bonfield2011,Rosario2012,Harris2016,Dong2016,Lanzuisi2017};
at lower redshifts, this correlation has also been seen to extend to
lower AGN luminosities \citep[e.g.][]{Rosario2012}, in line with the
strong correlation in SDSS found by \citet{Netzer2009}.

Following earlier investigations, we describe the dependence of the
host galaxy SFR on luminosity and redshift (for these powerful
quasars) as

\begin{equation}
  {\rm SFR} \propto L_{\rm bol}^\alpha (1+z)^\beta
\end{equation}

\begin{figure}
    \centering
    \includegraphics[width=\columnwidth]{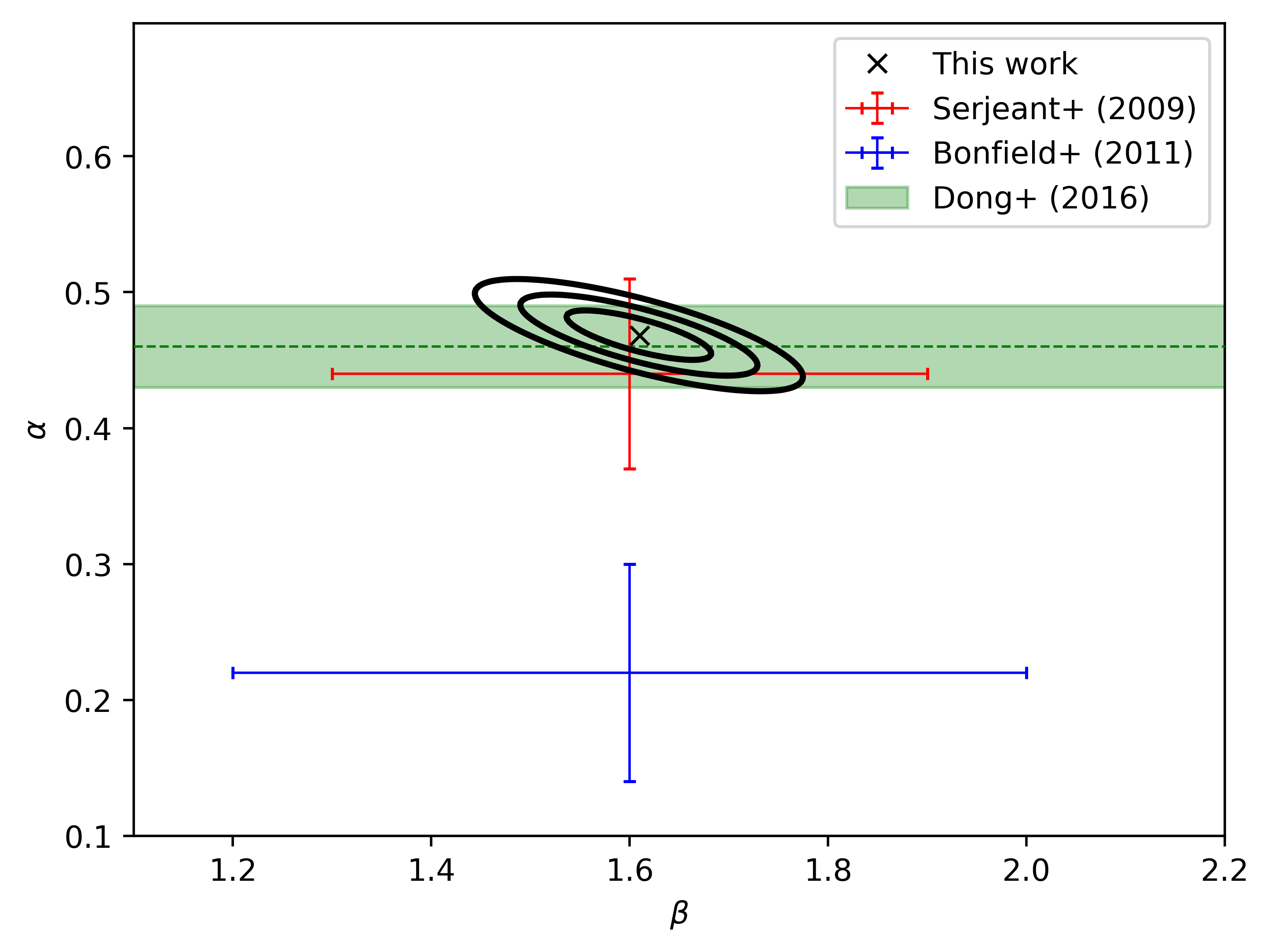}
    \caption{Constraints on the parameters $\alpha$ and $\beta$,
      relating star formation in quasar host galaxies to the redshift
      and luminosity of the quasar according to ${\rm SFR} \propto
      L_{\rm bol}^\alpha (1+z)^\beta$. Contours show the 1-, 2-,
      3-sigma confidence intervals. Large crosses show the derived
      values (and uncertainties) from \citet{Serjeant2009} and
      \citet{Bonfield2011} from far-infrared data, and the horizontal
      green shading shows the constraints on $\alpha$ derived by \citet{Dong2016}.
      \label{fig:alpha-beta}}
\end{figure}

\noindent Using the best-fit SFRs derived for each subsample in
redshift-luminosity space, we find constraints on $\alpha$ and $\beta$
as shown in Fig.~\ref{fig:alpha-beta}. As is evident from the upper
left panel of Fig.~\ref{fig:KS_grid_marg_fix}, which shows that the
SFR--$L_{\rm bol}$ relation is not strictly linear, the fitting
function does not provide a particularly good fit to the data across
all parameter space (reduced chi-squared $\sim 7$, dominated by the
lower redshift lowest luminosity points). However, the fits work well
at $z \ge 1$ and allow a like-for-like comparison of our results
against previous studies. Fig.~\ref{fig:alpha-beta} includes the
results obtained by \citet{Serjeant2009} and by
\citet{Bonfield2011}. In both cases these previous studies used
far-infrared estimates of the SFR, on the assumption that the far-IR
luminosity is dominated by the SFR contribution. As can be seen, our
results agree very well with those of Serjeant \& Hatziminaoglou, and
our value of $\beta = 1.61 \pm 0.05$ agrees with that of Bonfield
et~al.; their $\alpha$ is marginally discrepent, probably due to the
relatively small area of sky included in their study. A more recent
Herschel-based study by \citet{Dong2016} (also shown on
Fig.~\ref{fig:alpha-beta}) found $\alpha = 0.46 \pm 0.03$, in very
good agreement with our derived value of $\alpha = 0.47 \pm 0.01$. It
is also notable the scatter around the relation found by Dong \& Wu is
a few tenths of a dex, consistent with the value of $\sigma_{\Psi} =
0.42$\,dex found in our model for the width of the distribution of
star formation rates at given luminosity and redshift.  Overall, the
close agreement between our radio-derived SF properties of the quasar
hosts and those derived from far-IR data gives confidence that the
weak radio emission does indeed arise from star-formation, and is not
dominated by other processes such as disk coronal activity
\citep[see][]{Panessa2019}, which could also be correlated with the
quasar luminosity \citep[e.g.][]{Laor2008}.

\begin{figure}
    \centering
    \includegraphics[width=\columnwidth]{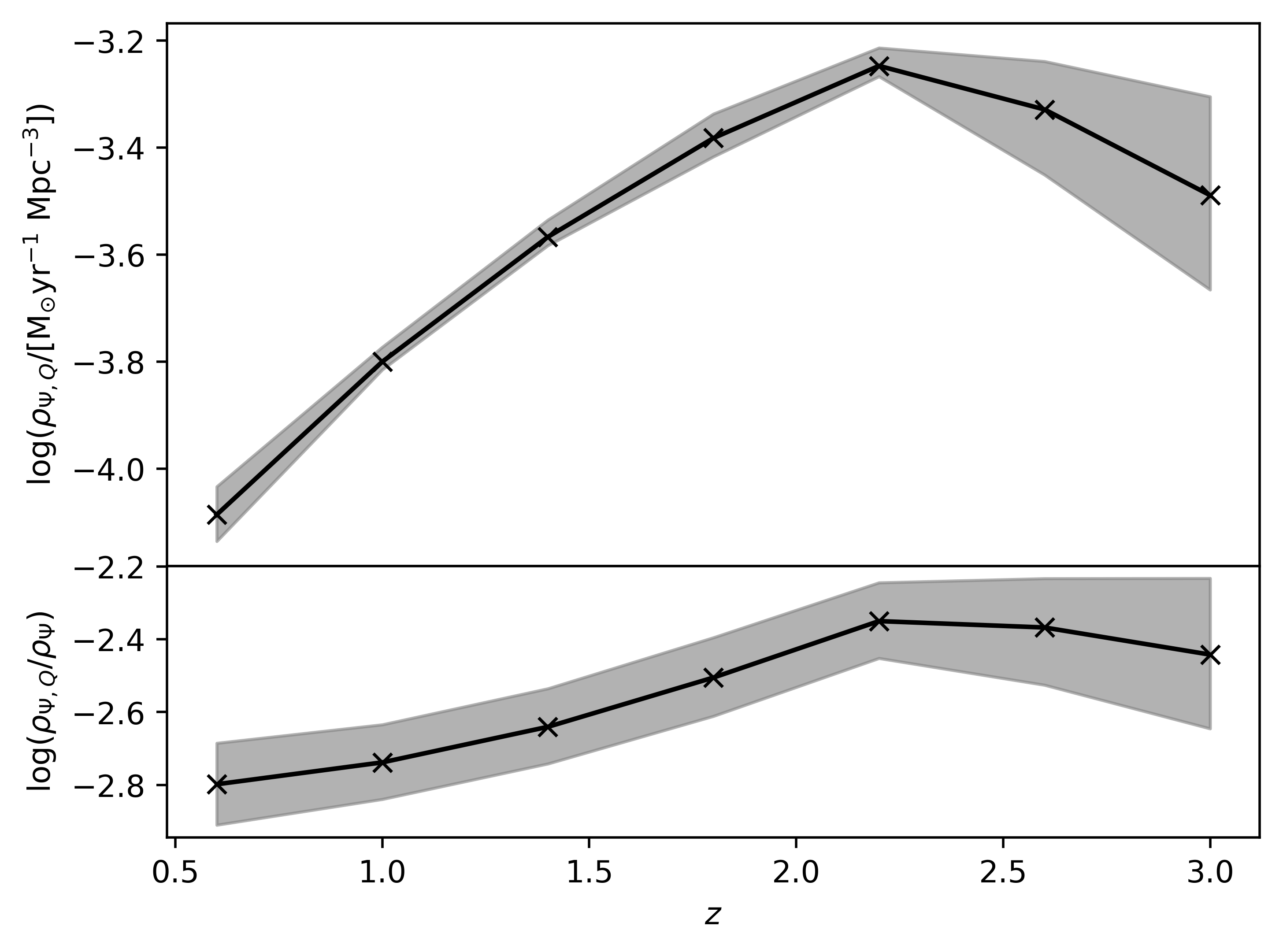}
    \caption{Top: the contribution of quasar host galaxies to the
      cosmic star-formation rate density, derived by combining the
      quasar luminosity function of \citet{Ross2013} with the
      relations between SFR and $M_i$ derived in
      Fig.~\ref{fig:KS_grid_marg_fix}. Bottom: the fractional
      contribution of quasars to the total cosmic star-formation rate
      density. In both panels, the grey-shaded region represents the
      uncertainty.
      \label{fig:cosmic_sfrd}}
\end{figure}

The dependence of the SFR on quasar luminosity at a given redshift can
be combined with the quasar luminosity function to investigate the
overall contribution of quasar host galaxies to the cosmic
star-formation rate density. We fit a linear function to the
SFR--$L_{\rm bol}$ relation in each redshift range, and combine this
with the quasar luminosity function derived by \citet{Ross2013}.
Specifically, we use Ross et~al.'s recommended double power-law model
with pure luminosity evolution over the redshift range $0.3<z<2.2$ and
luminosity and density evolution over the range $2.2 < z < 3.5$ (with
parameters derived from the Stripe 82 data). We integrate the
resultant luminosity function down to quasars with luminosity
$0.01L^*$ at each redshift. The results are shown in the upper panel
of Fig.~\ref{fig:cosmic_sfrd}, with statistical errors determined by
combining the uncertainties in our fitted SFR--$L_{\rm bol}$ relation
with the uncertainty in the faint-end slope derived by Ross et~al.:
these parameters dominate the statistical error budget. Systematic
errors may arise from the choice of integration limit, or from a
flattening of the SFR--$L_{\rm bol}$ relation at lower luminosities
\citep[e.g., see][]{Lanzuisi2017}, but these are more likely to shift
the whole distribution vertically than to change its shape.

The SFR density in quasar host galaxies is seen to increase with
increasing redshift from the current epoch back to $z \sim 2$, where
it flattens and then declines to higher redshifts.  This mirrors the
overall cosmic SFR density \citep[e.g.][]{Madau2014}. To compare
these, the lower panel of Fig.~\ref{fig:cosmic_sfrd} shows the
fractional contribution of quasar host galaxies to the cosmic SFR
density, derived by dividing the SFR density in quasars by the
functional fit to the total cosmic SFR density provided by
\citet{Madau2014}. Madau \& Dickinson integrate the cosmic SFR density
down to $0.03 L_{\rm SFR}^*$ at each redshift, which is broadly
comparable to our integration limit for the quasars, but we include an
additional 0.1 dex in quadrature in the uncertainty to account for
such systematic differences. Fig.~\ref{fig:cosmic_sfrd} shows that
quasar host galaxies account for approximately 0.15\% of all cosmic
star formation at $z \sim 0.5$, rising to 0.4\% at $z \sim 2$ and then
flattening towards higher redshifts. We discuss a possible explanation
for these results in the following subsection.

\subsection{Merger triggering of powerful quasar activity}
\label{Sec:disc_triggering}

Analysis of moderate luminosity AGN samples has indicated that they
lie close to the SFR versus stellar mass `main sequence' observed for
galaxies, with the increase in their SFR with redshift simply
mirroring the evolution of the star-forming main sequence \citep[see
  review by][]{HeckmanBest}. This, together with the lack of evidence
for a higher fraction of galaxy mergers or interactions in these AGN
compared to a control sample \citep[e.g.][]{Kocevski2012}, indicates
that the AGN activity in these objects is triggered by internal
secular processes \citep[see also][and references
    therein]{Smethurst2019}. This is consistent with the observation
that the host galaxies of moderate luminosity AGN often possess
`pseudo-bulges' \citep[e.g.][]{Capetti2006}; these are rapidly
rotating bulges, with power-law profiles and disky isophotes, which
are believed to form through secular processes such as bars and disk
instabilities in spiral galaxies \citep[see][]{Kormendy2004}.
\citet{Diamond2012} argue that in such AGN, the SFR on nuclear scales
(radius $<$ 1\,kpc) correlates well with the black hole accretion
rate, as both trace the very central gas densities of the system, but
the extended star-formation rates (which is all that can be measured
in high-redshift systems) do not correlate strongly; indeed any
residual correlation between global SFR and AGN luminosity may be
associated with both properties having a mutual dependence on galaxy
mass \citep[e.g.][]{Stemo2019}.

In contrast, at the highest quasar luminosities ($L_{\rm bol} \gta
10^{38}$\,W), and especially towards lower redshifts, the observed
SFRs of the quasar hosts lie significantly above the star-forming main
sequence, indicating that these objects are associated with starburst
activity. It has been widely argued that major galaxy mergers are the
most likely origin of both the starburst and the associated powerful
quasar \citep[e.g.][]{Sanders1988,Hopkins2006}; these objects are
typically hosted by very massive ellipticals, regardless of whether
they are radio-loud or radio-quiet
\citep[e.g.][]{Dunlop2003,Pagani2003}. Some observations
also find direct evidence of a high merger fraction at these high
luminosities \citep[e.g.][]{Treister2012,Goulding2018}; 
  this is particularly the case for the luminous reddened quasars
  \citep{Glikman2015}, with some authors suggesting that these are an
  evolutionary phase during which the dust associated with the merger
  starburst is blown out of the galaxy, before unreddened quasars are
  observed \citep[e.g.][]{Calistro2020}. It must be noted, however,
  that other authors have argued that the role of major mergers is
  sub-dominant \citep[][]{Hewlett2017} or unimportant
  \citep[][]{Marian2019} even at the highest luminosities, and
  therefore the requirement for major mergers to trigger the most
  luminous quasars remains controversial. If these objects are
  triggered by major mergers though, then the available gas mass and
the dynamical time of the system would influence both the SFR and the
black hole accretion rate, leading naturally to a correlation between
these two properties, as we observe in our study.

\citet{Lamastra2013} argue that starburst activity becomes
increasingly important (relative to quiescent star-formation) at
earlier cosmic times: they find that the fraction of the total cosmic
SFR density that is associated with starbursts increases by a factor
of 4 between $z\approx 0.1$ and $z \approx 5$. Similarly,
\citet{Martin2017} find a factor-of-two increase from $z=0$ to $z \sim
2$ in the fraction of cosmic SF that is associated with merging
systems. From a theoretical standpoint, the simulations of
\citet{Hopkins2010} also suggest that merger-induced starbursts
contribute 1-5\% of all star formation at $z \sim 0$, rising to 4-10\%
at $z>1$ and flattening at higher redshift. In
Fig.~\ref{fig:cosmic_sfrd} we found that the fraction of star
formation in powerful quasar host galaxies increased by a factor 2--3
between $z=0.5$ and $z \sim 2$ and then flattened; this mirrors the
trends found for starbursts and mergers, as would be expected for the
picture where powerful quasars are triggered by galaxy mergers.

Finally, in Section~\ref{Sec:BHmass_analysis} we considered the
influence of black hole mass on the results that we find. We showed in
Fig.~\ref{fig:KS_grid_marg_fix_bh} that for the bulk of the population
at $z \lta 2$, the SFR - AGN luminosity relation is the same in both
higher and lower black hole mass bins. This is to be expected if these
systems are driven by major mergers, since the black hole itself does
not play a major role in determining how quickly gas will be funnelled
down on to it. At $z>2$, however, if the possible small
  increase in the SFR in higher black hole mass systems compared to
  lower black hole masses is indeed real, then this could be
understood if, at those redshifts, we are reaching a regime where the
gas fractions in massive galaxies are high enough that the SFRs and
optical luminosities that we observe can be achieved in galaxies lying
on the star-forming main sequence; in such galaxies, a higher black
hole mass is likely to be correlated with a higher stellar mass and
hence a higher SFR.

\subsection{Quasar lifetimes}
\label{Sec:disc:lifetimes}

Numerically, we can convert the quasar optical luminosity to an
estimated growth rate of the central SMBH, $\dot{M}_{\rm BH}$ (as in
Section~\ref{Sec:accretion}). Although the bolometric correction from
optical luminosity is more uncertain than that from e.g. X-rays,
relying on an empirically-derived correlation, it is sufficiently
accurate for our purposes.

In Figure~\ref{fig:KS_grid_marg_fix}, we find that the ratio of the
star-formation rate in the host galaxy to the growth rate of the SMBH
is typically $\Psi/\dot{M}_{\rm BH} \sim 40$ (varying from
$\sim20-100$ across different redshift and luminosity bins). This
ratio is an order of magnitude lower than the ratio of bulge mass to
BH mass in present-day bulges \citep[$M_{\rm bulge}/M_{\rm
    BH}\approx500-700$; e.g.][]{Marconi2003,Haring2004}, suggesting
that the phase of quasar activity must be an order of magnitude
shorter than the duration of the star-formation activity of the
galaxy. This result is also consistent (for a merger--triggered
scenario) with the ratio between the fraction of the cosmic SFR
density in quasar host galaxies ($\sim 0.4$\% at z$\sim 2$;
Fig.~\ref{fig:cosmic_sfrd}) and that in merger-induced starbursts
\citep[$\sim 5$\% at $z \sim 2$;][]{Hopkins2010}. In major mergers,
the peak starburst activity is understood to last for a few tens of
Myr \citep[e.g.][]{Genzel1998,Bernloehr1993,Mihos1994}; this then
implies quasar lifetimes consistent with the lower end of the range
typically suggested by observations; $t_Q\approx 10^6-10^8$ yr
\citep{Martini2004}. Furthermore, it is also interesting to note that
we observe lower values of $\Psi/\dot{M}_{\rm BH}$ at the highest
accretion rates, suggesting that the lifetimes of the most luminous
quasars may be shorter than those of lower luminosities. This is
reasonable, since the higher luminosity quasars will be consuming
their gas supply more quickly.

There is an important caveat to these conclusions: it may well be the
case that the peak of quasar activity does not correspond to the peak
of star-formation activity, particularly if these quasars are part of
a merger-driven evolutionary sequence from ultra-luminous infrared
galaxies (starbursts) to quasars \citep{Sanders1988}.
\citet{Wild2010} studied the growth of BHs in galactic bulges in which
strong bursts of SF have recently occurred. They find that the black
hole growth peaks around 300\,Myr after the burst in
star-formation. They propose that BH growth has been driven primarily
by slow stellar ejecta from intermediate mass stars
\citep[cf.][]{Norman1988}, and that at earlier times the black hole
growth is suppressed by supernovae feedback. These effects would mean
that the currently estimated value of $\Psi$ may under-estimate the
star formation that occurs during the whole burst, and would then
permit longer quasar lifetimes. The quasar luminosity may also vary
over the quasar lifetime. However, any such variations are unlikely to
affect the qualitative conclusion that quasar lifetimes are shorter
than the period of star-formation activity, and that the most luminous
quasars have the shortest lifetimes.

\subsection{The powering of quasar jets}
\label{Sec:disc_jets}

No consistent discernible trend with redshift is observed for the
normalisation of the jet power, $f$. We do see hints of a decrease in
$f$ with increasing redshift for the highest optical luminosities (see
Fig.~\ref{fig:KS_grid_marg_fix}), in agreement with some previous
studies \citep[e.g.][]{Jiang2007,Balokovic2012,Kratzer2014}; this is
where the radio sources tend to be the most luminous and extended, and
might therefore be due to the increasing importance of inverse Compton
scattering losses in these sources towards higher redshifts, as
suggested by \citet{Gurkan2019}. Overall, however, the lack of strong
redshift dependence suggests that the physical properties that govern
the power of the radio jet are probably local properties of the
system. In agreement with the aforementioned studies, we do find
strong evidence of an increase in the fraction of sources at high
radio luminosities with increasing optical luminosity (or black hole
accretion rate): we find that $f$ increases by just over an order of
magnitude as the optical luminosity brightens by 4 magnitudes,
corresponding to $f \propto L_{\rm bol}^{\eta}$ with $\eta \sim
0.65$. This is broadly in line with the exponent of 0.85 suggested by
\citet{White2007} from analyses with the much shallower radio data
from the Faint Images of the Radio Sky at Twenty centimetres survey
\citep[FIRST;][]{Becker1995}. We emphasize this optical luminosity
dependence is a scaling of a full power-law distribution of jet powers
to typically higher powers: at all optical luminosities a large range
of jet powers is seen.

We also find weak evidence that the fraction of sources at high radio
luminosities may increase with an increase in the mass of the central
SMBH, at least out to $z=2$. Dependencies of the radio luminosity, or
RL fraction, on the BH mass have been argued by previous studies
\citep[e.g.][]{Laor2000,Lacy2001,Dunlop2003,McLureJarvis2004,Best2005}. It
is possible that any such dependence of jet power on black hole mass
could be due to a residual correlation with other properties, such as
stellar mass of the host. \citet{Sabater2019} found (albeit for
radiatively-inefficient AGN but the argument is the same) that
although the fraction of radio AGN increases with increasing BH mass,
once they disentangled the correlation between BH and stellar mass,
they found that the fraction of radio AGN was mainly driven by the
stellar mass. As we do not have information about stellar masses for
our sample, it is difficult to draw any direct conclusions as to which
property most directly drives any increase in jet power.

Our results are unable to provide direct evidence for the physical
mechanisms that produce quasar jets, however, they do allow us to
speculate. First, as discussed in Section~\ref{Sec:disc_bimodal}, our
model suggests that radio jets are ubiquitous in quasars -- or at very
least they exist down well into the traditional `radio-quiet' regime
where the jet luminosity is comparable to the starburst luminosity
\citep[or other sources of radio emission, cf.][]{Panessa2019}.  This
result is consistent with the detection of jet-like structures in deep
high-resolution observations of such quasars
\citep[e.g.][]{Hartley2019}. This suggests that the jet-launching
mechanism operates in all quasars, but with different powering
efficiency. Second, as discussed in Section~\ref{Sec:disc_SFRs}, we
interpret that the majority of the (high luminosity) quasars in our
sample are triggered by galaxy mergers. Third, we observe trends of
the jet power normalisation with optical luminosity (or SMBH growth
rate) and BH mass (or perhaps stellar mass).

One popular explanation for the varying power of radio jets is a
dependence on black hole spin \citep{Blandford1982}.  However, our
observations suggest a very wide range of jet powers, which would
require a correspondingly wide range of black hole spin parameters:
numerical models find it hard to produce this \citep{Volonteri2013}.
Furthermore, if the majority of our quasars (of high and low jet
powers) are indeed triggered by mergers, then since the orbits of two
colliding BHs would give rise to a significant amount of angular
momentum in the resulting BH, it would be surprising to find such a
large population of radio-quiet quasars. It therefore seems unlikely
that variations in black hole spin can be the main factor influencing
radio loudness.

An alternative hypothesis was put forward by \citet{Tchekhovskoy2011}
and developed by \citet{Sikora2013a} and \citet{Sikora2013b}: that the
main parameter driving the wide range of jet production efficiencies
is the magnetic flux threading a spinning black
hole. \citeauthor{Sikora2013b} argued that periods of hot accretion
are efficient at depositing magnetic flux close to the black
hole. Therefore, if a period of (cold-accretion) quasar activity had
been preceded by a period of hot accretion, as might happen for
example when a giant elliptical galaxy undergoes a merger with a disk
galaxy, then very powerful radio jets would be likely to result. We
have argued that the quasars in our sample are predominantly triggered
by major mergers, and hence in all cases these should have the
spinning black hole required. These mergers will have involved a mix
of progenitor galaxies (disk-elliptical and disk-disk mergers) and so
the black holes may be threaded by a wide variety of magnetic fluxes,
producing a significant range of jet efficiencies. Furthermore, as
well as high power radio jets produced in these `magnetically-choked
accretion flows', \citet{Sikora2013b} predict that low power or
intermittent jets in quasars will arise from fluctuating magnetic
fields arising in the corona above a thin accretion disk, or in a hot
inner region of an accretion flow. Combined, these could give rise to
the continuous distribution of jet powers that our model
adopts. Finally, for these accretion flows \citet{Sikora2013a} show
that (for fixed other parameters) the jet power increases with the
accretion rate (optical luminosity), broadly in line with the
correlation that we observe. Thus, our observations can all be well
explained by this model.

\section{Conclusions}
\label{Sec:conclusion}

We present a model of the radio luminosity distribution of quasars
that assumes that the radio emission of every quasar is a
superposition of two components: active galactic nuclei (jets) and
star-formation. We compare Monte Carlo simulated samples to our sample
of $\sim42,000$ quasars from the Sloan Digital Sky Survey quasar
catalogue fourteenth data release with the radio emission measured by
the LOFAR as part of the first data release of the LoTSS.

We find that our two-component model is valid in describing the
observed radio emission across a wide range of redshifts and optical
quasar luminosities. We therefore argue that an intrinsic bimodality
in the radio loudness distribution of quasars does not exist; instead,
`radio-loud' quasars are simply the luminous tail of a continuous jet
power distribution. Our analysis cannot prove whether or not quasar
jets are ubiquitous, but our results do suggest that the radio
emission from jets is contributing down at least to the level that
radio emission from SF becomes comparable. Our model naturally leads
to the expectation that some radio-quiet quasars will have their radio
emission dominated by small-scale jets, and others by star-formation,
in line with observations.

Given the validity of our model, we investigate how the parameters of
our model depend on redshift, optical luminosity (which we relate to
the SMBH growth rate) and BH mass. The width of the Gaussian SF
component and slope of the power-law AGN component are found not to
vary significantly so we fix the two parameters in our model. We find
a strong correlation between the mean SFR and the optical quasar
luminosity, with $SFR \propto L_{\rm bol}^{0.47 \pm 0.01}$. These
results are in line with other recent studies that probe the high
quasar luminosities characateristic of our sample, and with
far-infrared determinations. Unlike at lower quasar luminosities
(where the AGN activity is believed to be triggered by secular
processes and the host galaxies lie close to the star-forming main
sequence, leading to little correlation between star-formation and BH
acrretion rate), these high luminosity quasars are understood to be
triggered by massive galaxy mergers, where the gas fraction and
dynamical time of the system will influence both star formation and
black hole accretion, leading naturally to the observed
correlation. The ratio of the black hole growth rate to star formation
rate are observed to be an order of magnitude higher in these quasars
than the current black hole mass -- bulge mass ratio, implying that
quasar activity must be an intermittent phase.

We also investigate the cosmic star-formation history of quasar host
galaxies. We see an increase in the SFR of the quasar hosts out to
$z\sim2$, which then flattens in the range $z\sim2-3$. The integrated
star-formation rate density in quasar host galaxies contributes
roughly 0.15\% of the total cosmic star-formation rate density at $z
\sim 0.5$, increasing to around 0.4\% at $z \sim 2$ and then
flattening. This trend mirrors that of the importance of
merger-induced starbursts to cosmic star formation. We observe little
dependence of the SFR on BH mass until we get to the highest
redshifts, at which point we see weak evidence that increasing the BH
mass may correspond to a small increase in the SFR. These highest
redshift quasars lie closer to the star-forming main sequence and some
may be secularly triggered, in which case this observation would be
naturally explained by a correlation of both parameters with the
stellar mass of the host.

The normalisation of the jet power distribution is shown to have
little dependence on redshift, suggesting that the physical properties
responsible for producing powerful radio jets are local to the
system. We do observe an increase in the fraction of sources at high
radio luminosities with increasing optical luminosity (or BH growth
rate) and indications of an increase with BH mass, in line with
previous studies. Although our results do not allow a definitive
answer to be reached on the physical mechanisms that produce radio
jets, by considering the possible interpretations of our results, we
conclude that the model which can best explain our combination of
results is the one of \citet{Sikora2013b}, where the magnetic flux
threading the black hole is the primary factor influencing jet
production efficiency. This model is able to naturally produce the
very wide range of radio loudness required by observations, while also
giving a jet power to optical luminosity correlation.

We use our model to investigate the effectiveness of different
literature definitions of `radio loudness' for quasars. We find that
the traditional radio-loudness selection based on the ratio of
radio-to-optical luminosities, $R=L_{\rm 5GHz} / L_{\rm
  4400\text{\normalfont\AA}} > 10$, cleanly selects a sample of
jet-dominated sources ($L_{\rm AGN} / L_{\rm SF} > 1$), but does so in
a substantially incomplete manner: many quasars classed as
`radio-quiet' by this criteria have the majority of their radio
emission associated with the jet, which can be up to two orders of
magnitude brighter than that from star formation. We find that
definitions of radio-loudness based solely on radio luminosity perform
relatively poorly.
  
The potential of the Monte Carlo approach that we have adopted,
particularly if subsequently adapted to use a Bayesian framework, is
vast for this research. Wider areas of the LOFAR survey (Shimwell
et~al., in prep.), and new, much lower noise data in the LOFAR Deep
Fields \citep{Tasse2020,Sabater2020} can be seamlessly added to the
existing sample of quasars, to constrain the parameters of the model
further. With the larger samples, additional parameters can be
investigated, such as the quasar colour recently studied by
\citet{Klindt2019} and \citet{Rosario2020}, to help disentangle
evolutionary effects. Further work to disentangle the dependencies on
other relevant properties of the system such as stellar mass and the
Eddington ratio, would also be informative.

\section*{Acknowledgements}

PNB and JS are grateful for support from the UK STFC via grant
ST/R000972/1. MJJ acknowledges support from the UK STFC [ST/N000919/1]
and the Oxford Hintze Centre for Astrophysical Surveys which is funded
through generous support from the Hintze Family Charitable Foundation.
HR and KJD acknowledge support from the ERC Advanced Investigator
programme NewClusters 321271. IP acknowledges support from INAF under
the SKA/CTA PRIN `FORECaST' and the PRIN MAIN STREAM `SAuROS'
projects.  GCR acknowledges support from the Gruber Foundation/IAU
under the Gruber Foundation Fellowship. We thank the
  anonymous reviewer for helpful comments which improved the paper.

This paper is based (in part) on data obtained with the International
LOFAR Telescope (ILT) under project codes LC2\_038 and LC3\_008. LOFAR
\citep{vanHaarlem} is the Low Frequency Array designed and constructed
by ASTRON. It has observing, data processing, and data storage
facilities in several countries, that are owned by various parties
(each with their own funding sources), and that are collectively
operated by the ILT foundation under a joint scientific policy. The
ILT resources have benefitted from the following recent major funding
sources: CNRS-INSU, Observatoire de Paris and Universit{\'e} d'Orl{\'e}ans,
France; BMBF, MIWF-NRW, MPG, Germany; Science Foundation Ireland
(SFI), Department of Business, Enterprise and Innovation (DBEI),
Ireland; NWO, The Netherlands; The Science and Technology Facilities
Council, UK; Ministry of Science and Higher Education, Poland.

Funding for the Sloan Digital Sky Survey (I-IV) has been provided by
the Alfred P. Sloan Foundation, the Participating Institutions, the
National Science Foundation, the U.S. Department of Energy, the
National Aeronautics and Space Administration, the Japanese
Monbukagakusho, the Max Planck Society, and the Higher Education
Funding Council for England. The SDSS Web Site is
http://www.sdss.org/. The SDSS is managed by the Astrophysical
Research Consortium for the Participating Institutions.

\section*{Data Availability}

The datasets used in this paper were derived from sources in the
public domain: the LOFAR Two-Metre Sky Surveys (www.lofar-surveys.org)
and the Sloan Digital Sky Survey (www.sdss.org).

%%%%%%%%%%%%%%%%%%%%%%%%%%%%%%%%%%%%%%%%%%%%%%%%%%



\bibliographystyle{mnras}
\bibliography{biblio} % if your bibtex file is called example.bib

%%%%%%%%%%%%%%%%%%%%%%%%%%%%%%%%%%%%%%%%%%%%%%%%%%
%%%%%%%%%%%%%%%%% APPENDICES %%%%%%%%%%%%%%%%%%%%%

\appendix

\section{Threshold Tests}
\label{App:thresh}

\begin{figure*}
    \centering \includegraphics[width=15cm]{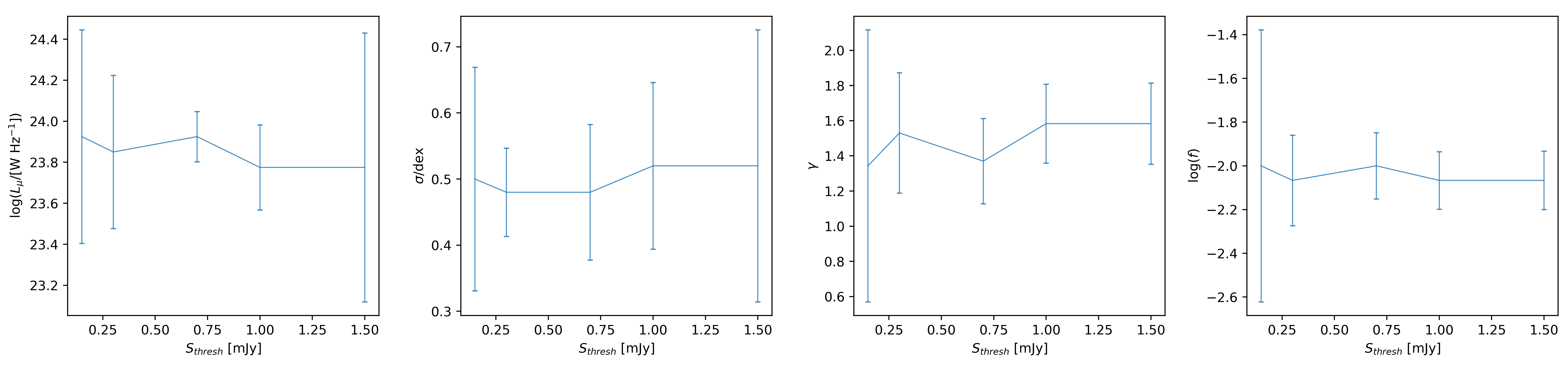}
    \caption{The four panels show how each of the best-fit values for
      the model parameters depends on the threshold flux density used
      to split the subsample in order to implement the method
      discussed in Section~\ref{Sec:ks}. The subsample shown is the
      same one as shown in Figure~\ref{fig:flux_example}. The plot
      demonstrates that the results obtained are robust against the
      choice of flux threshold to well within the errors; a similar
      conclusion is found for other subsamples.}
    \label{fig:thresh_params}
\end{figure*}

Since the definition of the threshold flux density is arbitrary, it is
important to test that our results do not depend on our choice of
$S_{\rm thresh}$. To do this, we computed the best-fit values of the
model parameters for given subsamples of $M_i - z$ space for a range
of threshold flux densities. The results for one such subsample are
shown in Fig.~\ref{fig:thresh_params}; similar results are found for
other subsamples. As can be seen from Fig.~\ref{fig:thresh_params},
the best-fit values for the parameters of the model are stable, within
the error bars, almost right down to a threshold flux density that
lies deep in the main quasar population at $S_{\rm thresh} =
0.15\mathrm{mJy}$. Therefore, a threshold flux density of $S_{\rm
  thresh} = 1\mathrm{mJy}$ should be suitable for our analyses.

%%%%%%%%%%%%%%%%%%%%%%%%%%%%%%%%%%%%%%%%%%%%%%%%%%

\section{Tables of best-fit values for model parameters}
\label{App:tabs}

\begin{table*}
    \caption{The variation of the best-fit values of the model
      parameters as a function of redshift and the optical luminosity
      of the quasar, for the full sample. Columns 1 and 2 indicate the
      redshift and optical luminosity bin under consideration. Column
      3 gives the black hole accretion rate corresponding to the
      optical luminosity (see Section~\ref{Sec:accretion}). Columns
      4-13 give the best-fit values for the model
        parameters, and their errors, as plotted in
        Figure~\ref{fig:KS_grid_collapse}: specifically the mean
        luminosity ($L_\mu$) associated with the star-forming
        component, the corresponding star-formation rate ($\Psi$), the
        width of the Gaussian distribution ($\sigma_\Psi =
        0.93\sigma$), the power-law slope of the AGN component
        ($\gamma$) and its normalisation term ($f$). $\Delta$ values
        indicate the relevant uncertainties. Columns 14-17 give the
        best-fit values when $\sigma$ and $\gamma$ are fixed
        (Section~\ref{Sec:fix}), as displayed in
        Figure~\ref{fig:KS_grid_marg_fix}.  \label{Tab:allsample}}
  \begin{tabular}{ccccccccccccccccc}
    \hline
    \multicolumn{3}{c}{\dots Input grid point \dots}  &
    \multicolumn{10}{c}{\dots\dots\dots\dots\dots\dots\dots Fitting results with four free parameters \dots\dots\dots\dots\dots\dots\dots} &
    \multicolumn{4}{c}{\dots Fixed $\sigma=0.45$, $\gamma = 1.4$ \dots} \\
    
    $z$ & $M_i$ & \hspace*{-1.5mm}$\mathrm{log}\dot{M}_{\rm BH}$\hspace*{-1.5mm} &
    \hspace*{-1.5mm}$\mathrm{log}L_{\mu}$\hspace*{-1.5mm} & \hspace*{-1.5mm}$\Delta \mathrm{log}L_{\mu}$\hspace*{-1.5mm} & $\mathrm{log}\Psi$ & $\Delta \mathrm{log}\Psi$ &  $\sigma_{\Psi}$ & $\Delta \sigma_{\Psi}$ & $\gamma$& $\Delta \gamma$ & $\mathrm{log}f$ & \hspace*{-1.5mm}$\Delta \mathrm{log}f$\hspace*{-1.5mm} & $\mathrm{log}\Psi_{\rm f}$ & $\Delta \mathrm{log}\Psi_{\rm f}$ & $\mathrm{log}f_{\rm f}$ & \hspace*{-1.5mm}$\Delta \mathrm{log}f_{\rm f}$\hspace*{-1.5mm} \\
    & \hspace*{-1.5mm}$(z=0)$\hspace*{-1.5mm} & \hspace*{-1.5mm}$[\mathrm{M}_{\odot}\mathrm{yr}^{-1}]$\hspace*{-1.5mm} & \hspace*{-1.5mm}$[\mathrm{W\ Hz}^{-1}]$\hspace*{-1.5mm} & \hspace*{-1.5mm}$[\mathrm{W\ Hz}^{-1}]$\hspace*{-1.5mm} & \hspace*{-2mm}$[\mathrm{M_{\odot}yr}^{-1}]$\hspace*{-2mm} & \hspace*{-1.5mm}$[\mathrm{M_{\odot}yr}^{-1}]$\hspace*{-1.5mm} & [dex] & [dex] & & & & & \hspace*{-2mm}$[\mathrm{M_{\odot}yr}^{-1}]$\hspace*{-2mm} &
    \hspace*{-1.5mm}$[\mathrm{M_{\odot}yr}^{-1}]$\hspace*{-1.5mm} & & \\
(1) & (2) & (3) & (4) & (5) & (6) & (7) & (8) & (9) & (10) & (11) & (12) & (13) & (14) & (15) & (16) & (17) \\
\hline
0.6 & -23.5 & -0.15 & 23.47 & 0.11 & 1.31 & 0.10 & 0.50 & 0.03 & 1.26 & 0.18 & -1.93 & 0.17 & 1.25 & 0.04 & -1.95 & 0.07 \\
0.6 & -22.5 & -0.53 & 23.10 & 0.06 & 0.96 & 0.05 & 0.48 & 0.03 & 1.29 & 0.05 & -2.13 & 0.11 & 0.93 & 0.02 & -2.15 & 0.04 \\
0.6 & -21.5 & -0.90 & 22.50 & 0.03 & 0.40 & 0.03 & 0.66 & 0.10 & 1.10 & 0.41 & -2.40 & 0.50 & 0.46 & 0.06 & -2.40 & 0.10 \\
1.0 & -24.5 &  0.22 & 23.85 & 0.09 & 1.66 & 0.08 & 0.48 & 0.04 & 1.21 & 0.13 & -1.60 & 0.07 & 1.56 & 0.01 & -1.65 & 0.02 \\
1.0 & -23.5 & -0.15 & 23.55 & 0.05 & 1.38 & 0.04 & 0.44 & 0.03 & 1.40 & 0.04 & -1.93 & 0.06 & 1.40 & 0.01 & -1.95 & 0.05 \\
1.0 & -22.5 & -0.53 & 23.32 & 0.01 & 1.17 & 0.01 & 0.39 & 0.04 & 1.32 & 0.04 & -2.13 & 0.07 & 1.14 & 0.01 & -2.25 & 0.05 \\
1.4 & -25.5 &  0.59 & 24.07 & 0.21 & 1.87 & 0.19 & 0.52 & 0.03 & 1.26 & 0.15 & -1.40 & 0.07 & 1.72 & 0.06 & -1.35 & 0.06 \\
1.4 & -24.5 &  0.22 & 23.77 & 0.15 & 1.59 & 0.14 & 0.44 & 0.04 & 1.48 & 0.08 & -1.73 & 0.04 & 1.66 & 0.01 & -1.70 & 0.06 \\
1.4 & -23.5 & -0.15 & 23.70 & 0.01 & 1.52 & 0.01 & 0.39 & 0.02 & 1.40 & 0.04 & -2.00 & 0.08 & 1.51 & 0.01 & -2.00 & 0.05 \\
1.4 & -22.5 & -0.53 & 23.55 & 0.02 & 1.38 & 0.02 & 0.33 & 0.03 & 1.32 & 0.05 & -2.07 & 0.12 & 1.30 & 0.05 & -2.10 & 0.15 \\
1.8 & -25.5 &  0.59 & 24.00 & 0.26 & 1.80 & 0.24 & 0.52 & 0.03 & 1.50 & 0.22 & -1.53 & 0.04 & 1.93 & 0.02 & -1.45 & 0.02 \\
1.8 & -24.5 &  0.22 & 23.92 & 0.05 & 1.73 & 0.04 & 0.44 & 0.02 & 1.50 & 0.03 & -1.93 & 0.05 & 1.82 & 0.05 & -1.90 & 0.11 \\
1.8 & -23.5 & -0.15 & 23.70 & 0.09 & 1.52 & 0.08 & 0.37 & 0.06 & 1.69 & 0.13 & -2.33 & 0.10 & 1.61 & 0.04 & -2.10 & 0.07 \\
2.2 & -26.5 &  0.97 & 24.52 & 0.11 & 2.29 & 0.10 & 0.50 & 0.04 & 1.34 & 0.22 & -1.40 & 0.11 & 2.24 & 0.06 & -1.30 & 0.13 \\
2.2 & -25.5 &  0.59 & 23.77 & 0.23 & 1.59 & 0.21 & 0.53 & 0.09 & 1.64 & 0.08 & -1.53 & 0.03 & 2.03 & 0.01 & -1.45 & 0.02 \\
2.2 & -24.5 &  0.22 & 24.07 & 0.02 & 1.87 & 0.02 & 0.42 & 0.05 & 1.40 & 0.04 & -1.87 & 0.07 & 1.87 & 0.01 & -1.85 & 0.05 \\
2.2 & -23.5 & -0.15 & 23.77 & 0.21 & 1.59 & 0.19 & 0.48 & 0.11 & 1.58 & 0.22 & -2.07 & 0.12 & 1.72 & 0.05 & -1.90 & 0.12 \\
2.6 & -26.5 &  0.97 & 24.67 & 0.36 & 2.43 & 0.33 & 0.53 & 0.15 & 1.42 & 0.31 & -1.73 & 0.20 & 2.40 & 0.07 & -1.40 & 0.12 \\
2.6 & -25.5 &  0.59 & 24.22 & 0.08 & 2.01 & 0.07 & 0.48 & 0.04 & 1.48 & 0.11 & -1.60 & 0.07 & 2.08 & 0.03 & -1.60 & 0.08 \\
2.6 & -24.5 &  0.22 & 24.07 & 0.04 & 1.87 & 0.04 & 0.44 & 0.09 & 1.42 & 0.10 & -1.93 & 0.11 & 1.87 & 0.05 & -1.85 & 0.11 \\
3.0 & -25.5 &  0.59 & 24.15 & 0.26 & 1.94 & 0.24 & 0.50 & 0.12 & 1.58 & 0.15 & -1.67 & 0.11 & 2.14 & 0.05 & -1.65 & 0.10 \\
3.0 & -24.5 &  0.22 & 24.15 & 0.08 & 1.94 & 0.07 & 0.40 & 0.11 & 1.34 & 0.07 & -1.67 & 0.14 & 1.87 & 0.07 & -1.60 & 0.18 \\
\hline
  \end{tabular}
  \vspace*{-0.4cm}
\end{table*}

In this appendix, we provide tables of best-fit values for the model
parameters when fitted to the full sample (Table~\ref{Tab:allsample}),
and when split by black hole mass (Table~\ref{Tab:BHmass}).

\begin{table*}
    \caption{The variation of the best fit star-formation rate
      ($\Psi$) and jet power normalisation parameter ($f$) as a
      function of redshift and the optical luminosity of the quasar,
      split by black hole mass. Columns 1 and 2 indicate the redshift
      and optical luminosity bin under consideration.  Columns 3-6
      give the best-fit values for the two model parameters, and their
      errors, for black hole masses below the median value in each
      bin, while columns 7-10 give the equivalent values for black
      hole masses above the median value. The data in this table are plotted in
      Figure~\ref{fig:KS_grid_marg_fix_bh}. \label{Tab:BHmass}}
  \begin{tabular}{cccccccccc}
    \hline
    \multicolumn{2}{c}{\dots Grid input \dots} &
    \multicolumn{4}{c}{\dots \dots \dots $M_{\rm BH} < M_{\rm BH,med}$\dots \dots \dots } &
    \multicolumn{4}{c}{\dots \dots \dots $M_{\rm BH} > M_{\rm BH,med}$\dots \dots \dots } \\
    $z$ & $M_i$ &  $\mathrm{log}\Psi_<$ & $\Delta \mathrm{log}\Psi_<$ & $\mathrm{log}f_<$ & $\Delta \mathrm{log}f_<$ &
    $\mathrm{log}\Psi_>$ & $\Delta \mathrm{log}\Psi_>$ & $\mathrm{log}f_>$ & $\Delta \mathrm{log}f_>$ \\
    & $(z=0)$ & \hspace*{-1.5mm}$[\mathrm{M_{\odot}yr}^{-1}]$\hspace*{-1.5mm} &
    \hspace*{-1.5mm}$[\mathrm{M_{\odot}yr}^{-1}]$\hspace*{-1.5mm} & & &
    \hspace*{-1.5mm}$[\mathrm{M_{\odot}yr}^{-1}]$\hspace*{-1.5mm} &
    \hspace*{-1.5mm}$[\mathrm{M_{\odot}yr}^{-1}]$\hspace*{-1.5mm} & & \\ 
    (1) & (2) & (3) & (4) & (5) & (6) & (7) & (8) & (9) & (10) \\ 
    \hline
0.6 & -23.5 & 1.30 & 0.03 & -2.15 & 0.14 & 1.19 & 0.05 & -1.80 & 0.07 \\ 
0.6 & -22.5 & 0.88 & 0.06 & -2.20 & 0.09 & 0.93 & 0.06 & -2.00 & 0.18 \\ 
0.6 & -21.5 & 0.62 & 0.07 & -2.65 & 0.35 & 0.46 & 0.12 & -2.30 & 0.25 \\ 
1.0 & -24.5 & 1.61 & 0.06 & -1.80 & 0.08 & 1.56 & 0.08 & -1.55 & 0.08 \\ 
1.0 & -23.5 & 1.40 & 0.05 & -1.90 & 0.07 & 1.40 & 0.03 & -1.70 & 0.05 \\ 
1.0 & -22.5 & 1.14 & 0.03 & -2.35 & 0.12 & 1.14 & 0.04 & -2.10 & 0.10 \\ 
1.4 & -25.5 & 1.77 & 0.04 & -1.50 & 0.04 & 1.61 & 0.10 & -1.20 & 0.05 \\ 
1.4 & -24.5 & 1.72 & 0.03 & -1.75 & 0.08 & 1.66 & 0.03 & -1.50 & 0.03 \\ 
1.4 & -23.5 & 1.61 & 0.07 & -2.05 & 0.16 & 1.51 & 0.05 & -1.85 & 0.10 \\ 
1.4 & -22.5 & 1.30 & 0.08 & -2.10 & 0.39 & 1.25 & 0.14 & -2.10 & 0.30 \\ 
1.8 & -25.5 & 1.98 & 0.08 & -1.45 & 0.08 & 1.93 & 0.06 & -1.40 & 0.13 \\ 
1.8 & -24.5 & 1.87 & 0.06 & -1.85 & 0.11 & 1.77 & 0.07 & -1.55 & 0.12 \\ 
1.8 & -23.5 & 1.72 & 0.04 & -2.25 & 0.32 & 1.66 & 0.04 & -1.90 & 0.13 \\ 
2.2 & -26.5 & 2.14 & 0.07 & -1.25 & 0.07 & 2.34 & 0.07 & -1.40 & 0.12 \\ 
2.2 & -25.5 & 1.93 & 0.07 & -1.35 & 0.11 & 2.14 & 0.02 & -1.60 & 0.07 \\ 
2.2 & -24.5 & 1.82 & 0.06 & -1.80 & 0.16 & 1.93 & 0.03 & -1.85 & 0.09 \\ 
2.2 & -23.5 & 1.56 & 0.10 & -1.75 & 0.15 & 1.82 & 0.06 & -2.05 & 0.29 \\ 
2.6 & -26.5 & 2.19 & 0.06 & -1.25 & 0.11 & 2.45 & 0.02 & -1.40 & 0.10 \\ 
2.6 & -25.5 & 1.98 & 0.06 & -1.50 & 0.11 & 2.14 & 0.06 & -1.55 & 0.12 \\ 
2.6 & -24.5 & 1.77 & 0.07 & -1.90 & 0.19 & 1.93 & 0.02 & -1.85 & 0.08 \\ 
3.0 & -25.5 & 2.13 & 0.04 & -1.80 & 0.15 & 2.08 & 0.05 & -1.35 & 0.06 \\ 
3.0 & -24.5 & 1.72 & 0.12 & -1.65 & 0.17 & 1.93 & 0.12 & -1.65 & 0.21 \\ 
\hline
  \end{tabular}
\end{table*}

%%%%%%%%%%%%%%%%%%%%%%%%%%%%%%%%%%%%%%%%%%%%%%%%%%

\section{$\mathbf{M_{i} - z}$ Binning Tests}
\label{App:binning}

It is important to test that the observed SFR trend with redshift is
not due to selection effects related to the binning in $M_i - z$
space, coupled with the strong dependence of the SFR on $M_i$. Each
bin has a width of $\Delta M_i = 1$ and so, for a given slice of $M_i$
(e.g. $-24 < M_i < -23$ across all $z$), the correlation between $M_i$
and $z$ imprinted by selection effects means that higher redshift bins
within the slice may contain quasars which are typically more luminous
than those in lower redshift bins. In turn, this could give rise to an
apparent trend between SFR and $z$, arising solely due to binning
biases.

\begin{figure}
    \centering
    \includegraphics[width=\columnwidth]{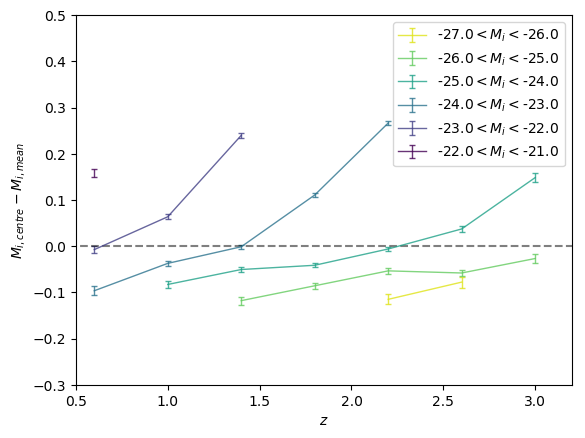}
    \caption{The distributions of the mean absolute \textit{i}-band
      magnitude, $M_{i,{\rm mean}}$, as a function of redshift for
      each slice of $M_i$. We offset $M_{i,{\rm mean}}$ with respect to the
      central absolute \textit{i}-band magnitude of the respective
      slice, $M_{i,{\rm centre}}$.}
    \label{fig:Mi_trend}
\end{figure}

To test this, the distribution of the mean absolute \textit{i}-band
magnitude, $M_{i,{\rm mean}}$, in each slice of $M_i$ was plotted as a
function of redshift, offset by the central absolute \textit{i}-band
magnitude of the slice. The results of this test can be seen in
Fig.~\ref{fig:Mi_trend}. We do observe a relatively small increase in
$M_{i,{\rm mean}}$, particularly at lower luminosities and especially
towards the upper end of each of their respective redshift
ranges. However, from Figure~\ref{fig:KS_grid_marg_fix}, such changes
in $M_i$ are only 0.2--0.3 magnitudes at most, which correspond to a
difference of only $\approx 0.1$\,dex in SFR; this is much smaller
that the variations seen in SFR with redshift. With this bias being
relatively small, and with evidence of the SFR in the range $-26 < M_i
< 25$ flattening off despite the bias at high $z$, we can infer the
derived results such as in Fig.~\ref{fig:KS_grid_marg_fix} are not
significantly affected by the bias.

% Don't change these lines
\bsp	% typesetting comment
\label{lastpage}
\end{document}